\documentclass[preprint,5p]{elsarticle}

\usepackage[english]{babel}

\usepackage{amsmath,amssymb}
\usepackage{booktabs}
\usepackage{amsthm}
\usepackage{array}
\usepackage{mathtools}
\usepackage[linesnumbered,longend,ruled,vlined]{algorithm2e}
\usepackage{pgfplots}
\usepackage{paralist}
\usepackage{url}
\usepackage{comment}

\usepackage{caption}
\usepackage{subcaption}

\usetikzlibrary{patterns}

\SetKwInOut{Input}{input}
\SetKwInOut{Output}{output}

\newcommand{\addcite}[1]{[ADDCITE: #1]}
\long\def\ignore#1{}

\newcommand{\skipfigure}[1]{\textcolor{green}{The figure is skipped to speed up compilation.}}
\renewcommand{\skipfigure}[1]{#1}

\newcommand{\tempclearpage}[0]{}
\renewcommand{\tempclearpage}[0]{\clearpage}
           
\usepackage{color}

\newcommand{\AJPii}[1]{{\leavevmode\color{red}[[AJP2: #1]]}}
\newcommand{\DKii}[1]{{\leavevmode\color{blue}[[DK2: #1]]}}

\let\AJPii=\ignore \let\DKii=\ignore

\usetikzlibrary{positioning,shapes,shadows,arrows}

\title{Markov Chain methods for the bipartite Boolean quadratic programming problem}

\renewcommand{\comma}{,\allowbreak\ }
\newcommand{\commanospace}{,\allowbreak}
\newcommand{\suchthat}{:\allowbreak\,}
\newcommand{\connect}[1]{\allowbreak\ \text{#1}\ \allowbreak{}}

\newcommand{\Msucc}[0]{\ensuremath{M^\text{succ}}}
\newcommand{\Mfail}[0]{\ensuremath{M^\text{fail}}}

{\par\bigskip\noindent\ignorespaces\begin{tabular*}{\textwidth}{@{} p{0.35\textwidth} @{\hspace{0.5em}} l} #1}%
{\end{tabular*}\\\smallskip\par\noindent\ignorespacesafterend}

\newcommand{\krow}[1]{CMCS[#1-row]}
\newcommand{\reduced}[1]{CMCS[2-row reduced]}
\newcommand{\opprob}[0]{Op.\ Prob.}

\newcommand{\component}[1]{\textsc{#1}}
\newcommand{\OptX}{\component{OptX}}
\newcommand{\OptY}{\component{OptY}}
\newcommand{\FlpX}{\component{FlpX}}
\newcommand{\FlpY}{\component{FlpY}}
\newcommand{\Repair}{\component{Repair}}
\newcommand{\MutX}[1]{\component{MutX#1}}
\newcommand{\MutY}[1]{\component{MutY#1}}

\def\makeclean{
	\let\AJP=\ignore 
	\let\DK=\ignore	
    \let\AJPii=\ignore 
	\let\DKii=\ignore
    \let\FILE=\ignore
    \let\tempclearpage=\relax
    \let\addcite=\ignore
}

\date{}

    \let\tempclearpage=\relax

\begin{document}

\author[essex,nott,sfu]{Daniel Karapetyan\corref{cor1}}
\ead{daniel.karapetyan@gmail.com}
\author[sfu]{Abraham P.~Punnen}
\ead{apunnen@sfu.ca}
\author[nott]{Andrew J.~Parkes}
\ead{andrew.parkes@nottingham.ac.uk}

\cortext[cor1]{Corresponding author.  Tel: +44\,1206\,872861}
\address[essex]{Institute for Analytics and Data Science, University of Essex, Colchester, CO4\,3SQ, UK}
\address[nott]{ASAP Research Group, School of Computer Science, University of Nottingham, Jubilee Campus, Wollaton Road, Nottingham, NG8\,1BB, UK}
\address[sfu]{Department of Mathematics, Simon Fraser University Surrey, Central City, 250-13450 102nd AV, Surrey, British Columbia, V3T\,0A3, Canada}

\begin{abstract}
 We study the Bipartite Boolean Quadratic Programming Problem (BBQP) which is an extension of the well known Boolean Quadratic Programming Problem (BQP).
 Applications of the BBQP include mining discrete patterns from binary data, approximating matrices by rank-one binary matrices, computing the cut-norm of a matrix, and solving optimisation problems such as maximum weight biclique, bipartite maximum weight cut, maximum weight induced sub-graph of a bipartite graph, etc.  
 For the BBQP, we first present several algorithmic components, specifically, hill climbers and mutations, and then show how to combine them in a high-performance metaheuristic.
 Instead of hand-tuning a standard metaheuristic to test the efficiency of the hybrid of the components, we chose to use an automated generation of a multi-component metaheuristic to save human time, and also improve objectivity in the analysis and comparisons of components.
 For this we designed a new metaheuristic schema which we call Conditional Markov Chain Search (CMCS).
 We show that CMCS is flexible enough to model several standard metaheuristics; this flexibility is controlled by multiple numeric parameters, and so is convenient for automated generation.
 We study the configurations revealed by our approach and show that the best of them outperforms the previous state-of-the-art BBQP algorithm by several orders of magnitude.
 In our experiments we use benchmark instances introduced in the preliminary version of this paper and described here, which have already become the de facto standard in the BBQP literature.

\end{abstract}

\maketitle

\section{Introduction}
\label{sec:introduction}

 The (Unconstrained) Boolean Quadratic Programming Problem (BQP) is to
\begin{align*}
\text{maximise } & f(x) = x^T Q' x + c' x + c'_0\\
\mbox{subject to } & x \in \{ 0, 1 \}^n,
\end{align*}
where $Q'$ is an $n \times n$ real matrix, $c'$ is a row vector in $\mathbb{R}^n$, and $c'_0$ is a constant.  
 The BQP is a well-studied problem in the operational research literature~\cite{Billionnet2004}.  
 The focus of this paper is on a problem closely related to BQP, called the \emph{Bipartite (Unconstrained) Boolean Quadratic Programming Problem} (BBQP)~\cite{Punnen2012}\@.  
 BBQP can be defined as follows:
\begin{align*}
\text{maximise } & f(x, y) = x^TQy + cx + dy + c_0\\
\text{subject to } & x \in \{ 0, 1 \}^m, y \in \{ 0, 1 \}^n,
\end{align*}
where $Q = (q_{ij})$ is an $m \times n$ real matrix, $c = (c_1, c_2, \ldots, c_m)$ is a row vector in $\mathbb{R}^m$, $d = (d_1, d_2, \ldots, d_n)$ is a row vector in $\mathbb{R}^n$, and $c_0$ is a constant.
 Without loss of generality, we assume that $c_0 = 0$, and $m \le n$ (which can be achieved by simply interchanging the rows and columns if needed).
 In what follows, we denote a BBQP instance built on matrix $Q$, row vectors $c$ and $d$ and $c_0 = 0$ as BBQP$(Q, c, d)$, and $(x, y)$ is a feasible solution of the BBQP if $x \in \{ 0, 1 \}^m$ and $y \in \{ 0, 1 \}^n$.  
 Also $x_i$ stands for the $i$th component of the vector $x$ and $y_j$ stands for the $j$th component of the vector $y$.


 A graph theoretic interpretation of the BBQP can be given as follows~\cite{Punnen2012}.  
 Let $I = \{ 1 \commanospace 2 \commanospace \ldots \commanospace m \}$ and $J = \{ 1 \commanospace 2 \commanospace \ldots \commanospace n \}$.  
 Consider a bipartite graph $G = (I, J, E)$.  
 For each node $i \in I$ and $j \in J$, respective costs $c_i$ and $d_j$ are prescribed.  
 Furthermore, for each $(i,j) \in E$, a cost $q_{ij}$ is given.  
 Then the \emph{Maximum Weight Induced Subgraph Problem} on $G$ is to find a subgraph $G' = (I', J', E')$ such that $\sum_{i \in I'} c_i + \sum_{j \in J'} d_j + \sum_{(i,j) \in E'} q_{ij}$ is maximised, where $I' \subseteq I$, $J' \subseteq J$ and $G'$ is induced by $I' \cup J'$.  
 The Maximum Weight Induced Subgraph Problem on $G$ is precisely the BBQP, where $q_{ij} = 0$ if $(i, j) \notin E$.

 There are some other well known combinatorial optimisation problems that can be modelled as a BBQP\@. 
 Consider the bipartite graph $G = (I, J, E)$ with $w_{ij}$ being the weight of the edge $(i, j) \in E$\@. 
 Then the \emph{Maximum Weight Biclique Problem} (MWBP)~\cite{Ambuhl2011,Tan2008} is to find a biclique in $G$ of maximum total edge-weight.  
 Define
\[
	q_{ij} =
	\begin{cases}
		w_{ij} &\mbox{if } (i,j) \in E, \\
		-M & \mbox{otherwise,}
	\end{cases}
\]
where $M$ is a large positive constant.
 Set $c$ and $d$ as zero vectors.  
 Then BBQP$(Q, c, d)$ solves the MWBP~\cite{Punnen2012}.
 This immediately shows that the BBQP is NP-hard and one can also establish some approximation hardness results with appropriate assumptions~\cite{Ambuhl2011,Tan2008}.  
 Note that the MWBP has applications in data mining, clustering and bioinformatics~\cite{Chang2012,Tanay2002} which in turn become applications of BBQP\@.

 Another application of BBQP arises in approximating a matrix by a rank-one binary matrix~\cite{Gillis2011,Koyuturk2005,Koyuturk2006,Lu2011,Shen2009}.
 For example, let $H = (h_{ij})$ be a given $m \times n$ matrix and we want to find an $m \times n$ matrix $A = (a_{ij})$, where $a_{ij} = u_i v_j$ and $u_i, v_j \in \{ 0, 1 \}$, such that $\sum_{i = 1}^m \sum_{j = 1}^n (h_{ij} - u_i v_j)^2$ is minimised.
 The matrix $A$ is called a rank one approximation of $H$ and can be identified by solving the BBQP with $q_{ij} = 1 - 2h_{ij}$, $c_i = 0$ and $d_j = 0$ for all $i \in I$ and $j \in J$.  
 Binary matrix factorisation is an important topic in mining discrete patterns in binary data~\cite{Lu2011,Shen2009}.
 If $u_i$ and $v_j$ are required to be in $ \{-1,1\}$ then also the resulting factorisation problem can be formulated as a BBQP.

 The Maximum Cut Problem on a bipartite graph (MaxCut) can be formulated as BBQP~\cite{Punnen2012} and this gives yet another application of the model.  
 BBQP can also be used to find approximations to the cut-norm of a matrix~\cite{Alon2006}.

 For theoretical analysis of approximation algorithms for BBQP, we refer to~\cite{Punnen2015}. 

 A preliminary version of this paper was made available to the research community in 2012~\cite{Karapetyan2012}.  
 Subsequently Glover et al.~\cite{Glover2015} and Duarte et al.~\cite{Duarte2014} studied heuristic algorithms for the problem.
 The testbed presented in our preliminary report~\cite{Karapetyan2012} continues to be the source of benchmark instances for the BBQP\@.
 In this paper, in addition to providing a detailed description of the benchmark instances, we refine the algorithms reported in~\cite{Karapetyan2012}, introduce a new class of algorithms and give a methodology for automated generation of a multi-component metaheuristic.  
 By (algorithmic) component we mean a black box algorithm that modifies the given solution.  
 All the algorithmic components can be roughly split into two categories: hill climbers, i.e.\ components that guarantee that the solution not be worsened, and mutations, i.e.\ components that usually worsen the solution.
 Our main goals are to verify that the proposed components are sufficient to build a high-performance heuristic for BBQP and also investigate the most promising combinations.
 By this computational study, we also further support the ideas in the areas of automated parameter tuning and algorithm configuration (e.g. see \cite{Belarmino2006,BezerraEtal2015:component-MOO,HutterEtal2009:ParamILS,Hutter2007}).
 Thus we rely entirely on automated configuration.  
 During configuration, we use smaller instances compared to those in our benchmark.
 This way we ensure that we do not over-train our metaheuristics to the benchmark instances -- an issue that is often quite hard to avoid with manual design and configuration.
 We apply the resulting multi-component metaheuristic to our benchmark instances demonstrating that a combination of several simple components can yield powerful metaheuristics clearly outperforming the state-of-the-art BBQP methods.

 The main contributions of the paper include:
 \begin{itemize}
 	\item 
    In Section~\ref{sec:components}, we describe several BBQP algorithmic components, one of which is completely new.

    \item
    In Section~\ref{sec:cmchh} we take the Markov Chain idea, such as in the Markov Chain Hyper-heuristic \cite{McClymontKeedwell:GECCO2011:MCHH-selective-HH}, but restrict it to use static weights (hence having no online learning, and so, arguably, not best labelled as a ``hyper-heuristic''), but instead adding a powerful extension to it, giving what we call ``Conditional Markov Chain Search (CMCS)''.

 	\item
    In Section~\ref{sec:testbed} we describe five classes of instances corresponding to various applications of BBQP\@.
    Based on these classes, a set of benchmark instances is developed.
    These test instances were first introduced in the preliminary version of this paper~\cite{Karapetyan2012} and since then used in a number of papers~\cite{Duarte2014,Glover2015} becoming de facto standard testbed for the BBQP.
    
    \item
    In Section~\ref{sec:tuning} we use automated configuration of CMCS to demonstrate the performance of individual components and their combinations, and give details sufficient to reproduce all of the generated metaheuristics. 
    We also show that a special case of CMCS that we proposed significantly outperforms several standard metaheuristics, on this problem.
    
    \item
    In Section~\ref{sec:evaluation} we show that our best machine-generated metaheuristic is, by several orders of magnitude, faster than the previous state-of-the-art BBQP method.
        
 \end{itemize}

\tempclearpage
\section{Algorithmic Components}
\label{sec:components}

 In this section we introduce several algorithmic components for BBQP\@.
 Except for `\Repair{}' and `Mutation-X/Y', these components were introduced in~\cite{Karapetyan2012}.
 A summary of the components discussed below is provided in Table~\ref{tab:components}.
 The components are selected to cover a reasonable mix of fast and slow hill climbing operators for intensification, along with mutation operators that can be expected to increase diversification, and with \Repair{} that does a bit of both.
 Note that a hill climbing component can potentially implement either a simple improvement move or a repetitive local search procedure with iterated operators that terminates only when a local maximum is reached.
 However in this project we opted for single moves leaving the control to the metaheuristic framework.

\begin{table*}[htb]
\begin{tabular}{@{} ll @{}}
\toprule
Name & Description \\
\midrule
\multicolumn{2}{@{}l}{\textbf{--- Hill climbing operators: that is, components guaranteeing that the solution will not be worsened}} \\
\OptX{} & \component{Optimise-X}, Section~\ref{sec:optimisexy}.  Fixes vector $y$ while optimising $x$. \\
\OptY{} & As \OptX{}, but reversing roles of $x$ and $y$. \\
\FlpX{} & \component{Flip-X}, Section~\ref{sec:flip}.  Checks if flipping $x_i$ for some $i \in I$ and subsequently optimising $y$ improves the solution. \\
\FlpY{} & As \FlpX{}, but reversing roles of $x$ and $y$.  \\[2ex]
\multicolumn{2}{@{}l}{\textbf{--- Mutations: that is, components that may worsen the solution}} \\
\Repair{} & \Repair{}, Section~\ref{sec:repair}.  Finds a single term of the objective function that can be improved and ``repairs'' it. \\
\MutX{4} & Mutation-X(4), Section~\ref{sec:mutation}.  Flips $x_i$ for four randomly picked $i \in I$. \\
\MutY{4} & As \MutX{4}, but reversing roles of $x$ and $y$. \\
\MutX{16} & As \MutX{4}, but for 16 randomly picked $x_i$. \\
\MutY{16} & As \MutY{4}, but for 16 randomly picked $y_i$. \\
\bottomrule
\end{tabular}
\caption{
	List of the algorithmic components used in this paper, and described in Section~\ref{sec:components}
}
\label{tab:components}
\end{table*}  
 
\subsection{Components: \component{Optimise-X} / \component{Optimise-Y}}
\label{sec:optimisexy}
  
 Observe that, given a fixed vector $x$, we can efficiently compute an optimal $y = y_\text{opt}(x)$:
\begin{equation}
	\label{eq:optimal-y}
	y_\text{opt}(x)_j = \begin{cases}
		1 & \text{if } \displaystyle{\sum_{i \in I} q_{ij} x_i + d_j > 0}, \\
		0 & \text{otherwise.}
	\end{cases}
\end{equation}
 This suggests a hill climber operator \emph{\component{Optimise-Y}} (\OptY{}) that fixes $x$ and replaces $y$ with $y_\text{opt}(x)$.
 Equation (\ref{eq:optimal-y}) was first introduced in~\cite{Punnen2012} and then used as a neighbourhood search operator in \cite{Karapetyan2012}, \cite{Duarte2014} and \cite{Glover2015}.
  
 \OptY{} implements a hill climber operator in the neighbourhood $N_\text{\OptY{}}(x, y) = \{ (x, y') \suchthat y' \in \{ 0, 1 \}^n \}$, where $(x, y)$ is the original solution.
 Observe that the running time of \OptY{} is polynomial and the size of the neighbourhood $|N_\text{\OptY{}}(x, y)| = 2^n$ is exponential; hence \OptY{} corresponds to an operator that could be used in a very large-scale neighbourhood search (VLNS), a method that is often considered as a powerful approach to hard combinatorial optimisation problems~\cite{Ajuja2002}.
  
 Observe that \OptY{} finds a local maximum after the first application because $N(x, y) = N(x, y_\text{opt}(y))$ (that is, it is an ``idempotent operator''); hence, there is no gain from applying \OptY{} again immediately after it was applied.
 Though, for example, iterating and alternating between between \OptX{} and \OptY{} would give a VLNS.

 Note that $y_\text{opt}(x)_j$ can take any value if $\sum_{i \in I} q_{ij} x_i + d_j = 0$, without affecting the objective value of the solution.
 Thus, one can implement various ``tie breaking'' strategies including randomised decision whether to assign 0 or 1 to $y_\text{opt}(x)_j$, however in that case \OptY{} would become non-deterministic.
 In our implementation of \OptY{} we preserve the previous value by setting $y_\text{opt}(x)_j = y_j$ for every $j$ such that $\sum_{i \in I} q_{ij} x_i + d_j = 0$.
 As will be explained in Section~\ref{sec:solution-representation}, changing a value $y_j$ is a relatively expensive operation and thus, whenever not necessary, we prefer to avoid such a change.
 
 By interchanging the roles of rows and columns, we also define
\begin{equation}
	\label{eq:optimal-x}
	x_\text{opt}(y)_i = \begin{cases}
		1 & \text{if } \displaystyle{\sum_{j \in J} q_{ij} y_j + c_i > 0}, \\
		0 & \text{otherwise,}
	\end{cases}
\end{equation}
and a hill climber operator \emph{\component{Optimise-X}} (\OptX{}) with properties similar to those of \OptY{}.

\subsection{Components: \component{Flip-X} / \component{Flip-Y}}
\label{sec:flip}

 This class of components is a generalisation of the previous one.
 In \emph{\component{Flip-X}} (\FlpX{}), we try to flip $x_i$ for every $i \in I$, each time re-optimising $y$.
 More formally, for $i = 1, 2, \ldots, m$, we compute $x' = (x_1 \comma \ldots \comma x_{i-1} \comma 1 - x_i \comma x_{i + 1} \comma \ldots \comma x_m)$ and then verify if solution $(x', y_\text{opt}(x'))$ is an improvement over $(x, y)$.
 Each improvement is immediately accepted, but the search carries on for the remaining values of $i$.
 In fact, one could consider a generalisation of \component{Flip-X} that flips $x_i$ for several $i$ at a time.
 However, exploration of such a neighbourhood would be significantly slower, and so we have not studied such a generalisation in this paper.
 
 By row/column interchange, we also introduce the \emph{\component{Flip-Y}} (\FlpY{}) hill climbing operator.
 Clearly, \FlpX{} and \FlpY{} are also VLNS operators, though unlike \OptX{} and \OptY{} they are not idempotent and so could be used consecutively.

 \FlpX{} and \FlpY{} were first proposed in~\cite{Punnen2012} and then used in \cite{Glover2015}.

\subsection{Components: \Repair{}}
\label{sec:repair}

 While all the above methods were handling entire rows or columns, \emph{\Repair{}} is  designed to work on the level of a single element of matrix $Q$.
 \Repair{} is a new component inspired by the WalkSAT heuristic for SAT problem \cite{Papadimitriou1991:selecting-truth-assignment,Selman95localsearch} in that it is a version of `iterative repair' \cite{ZwebenEtal1993:scheduling-with-iterative-repair} that tries to repair some significant `flaw' (deficiency of the solution) even if this results in creation of other flaws, in a hope that the newly created flaws could be repaired later.
 This behaviour, of forcing the repair of randomly selected flaws, gives some stochasticity to the search that is also intended to help in escaping from local optima.
 
 Recall that the objective value of BBQP includes terms $q_{ij} x_i y_j$.
 For a pair $(i, j)$, there are two possible kinds of flaws: either $q_{ij}$ is negative but is included in the objective value (i.e.\ $x_i y_j = 1$), or it is positive and not included in the objective value (i.e.\ $x_i y_j = 0$).
 The \Repair{} method looks for such flaws, especially for those with large $|q_{ij}|$.
 For this, it uses the tournament principle; it randomly samples pairs $(i, j)$ and picks the one that maximises $(1 - 2 x_i y_j) q_{ij}$.
 Once an appropriate pair $(i, j)$ is selected, it `repairs' the flaw; if $q_{ij}$ is positive then it sets $x_i = y_j = 1$; if $q_{ij}$ is negative then it sets either $x_i = 0$ or $y_j = 0$ (among these two options it picks the one that maximises the overall objective value).
 Our implementation of \Repair{} terminates after the earliest of two: (i) finding 10 flaws and repairing the biggest of them, or (ii) sampling 100 pairs $(i, j)$.
 
 Note that one could separate the two kinds of flaws, and so have two different methods: \component{Repair-Positive}, that looks for and repairs only positive `missing' terms of the objective function, and \component{Repair-Negative}, that looks for and repairs only negative included terms of the objective function.
 However, we leave these options to future research.

\subsection{Components: \component{Mutation-X} / \component{Mutation-Y}}
\label{sec:mutation}
 
 In our empirical study, we will use some pure mutation operators of various strengths to escape local maxima.
 For this, we use the $N_\text{\OptX{}}(x, y)$ neighbourhood.
 Our \emph{\component{Mutation-X($k$)}} operator picks $k$ distinct $x$ variables at random and then flips their values, keeping $y$ unchanged.
 Similarly we introduce \emph{\component{Mutation-Y($k$)}}.
 In this paper we use $k \in \{ 4, 16 \}$, and so have components which we call \MutX{4}, \MutX{16}, \MutY{4} and \MutY{16}.
 
 An operator similar to Mutation-X/Y was used in~\cite{Duarte2014}. 

\tempclearpage
\section{The Markov Chain Methods}
\label{sec:cmchh}

 The algorithmic components described in Section~\ref{sec:components} are designed to work within a metaheuristic; analysis of each component on its own would not be sufficient to conclude on its usefulness within the context of a multi-component system.
 To avoid bias due to picking one or another metaheuristic, and to save human time on hand-tuning it, we chose to use a generic schema coupled with automated configuration of it. 
 
\subsection{Conditional Markov Chain Search (CMCS)}
 
 The existing framework that was closest to our needs was the Markov Chain Hyper-Heuristic (MCHH) \cite{McClymontKeedwell:GECCO2011:MCHH-selective-HH}.
 MCHH is a relatively simple algorithm that applies components in a sequence.
 This sequence is a Markov chain; the `state' in the Markov chain is just the operator that is to be applied, and so the Markov nature means that the transition to a new state (component/operator) only depends on the currently-applied component and transition probabilities.
 Transition probabilities, organised in a transition matrix, are obtained in MCHH dynamically, by learning most successful sequences.
 
 While MCHH is a successful approach capable of effectively utilising several algorithmic components, it does not necessarily provide the required convenience of interpretation of performance of individual components and their combinations because the transition probabilities in MCHH change dynamically.
 To address this issue, we chose to fix the transition matrix and learn it offline.
 We can then perform the analysis by studying the learnt transition probabilities.
  
 The drawback of learning the probabilities offline is that MCHH with static transition matrix receives no feedback from the search process and, thus, has no ability to respond to the instance and solution properties.
 To enable such a feedback, we propose to extend the state of the Markov chain with the information about the outcome of the last component execution; this extension is simple but will prove to be effective.
 In particular, we suggest to distinguish executions that improved the solution quality, and executions that worsened, or did not change, the solution quality.

 We call our new approach \emph{Conditional Markov Chain Search} (CMCS)\@.
 It is parameterised with two transition matrices: $\Msucc$ for transitions if the last component execution was successful (improved the solution), and $\Mfail$ for transitions if the last component execution failed (has not improved the solution).\footnote{
 	Note that executions that do not change the solution quality at all are also considered as a failure.
    This allows us to model a hill climber that is applied repeatedly until it becomes trapped in a local maximum.

 Let $\mathcal{H}$ be the pool of algorithmic components.
 CMCS is a single-point metaheuristic that applies one component $h \in \mathcal{H}$ at a time, accepting both improving and worsening moves.
 The next component $h' \in \mathcal{H}$ to be executed is determined by a function $\mathit{next} : \mathcal{H} \rightarrow \mathcal{H}$. 
 In particular, $h'$ is chosen using roulette wheel with probabilities $p_{hh'}$ of transition from $h$ to $h'$ defined by matrix $\Msucc$ if the last execution of $h$ was successful and $\Mfail$ otherwise.
 All the moves are always accepted in CMCS\@. 
 Pseudo-code of the CMCS schema is given in Algorithm~\ref{alg:cmchh}.
 }

\begin{algorithm}[htb]
\Input{Ordered set of components $\mathcal{H}$;}
\Input{Matrices $\Msucc$ and $\Mfail$ of size $|\mathcal{H}| \times |\mathcal{H}|$;}
\Input{Objective function $f(S)$ to be maximised;}
\Input{Initial solution $S$;}
\Input{Termination time $\mathit{terminate\text{-}at}$;}

$S^* \gets S$\;
$h \gets 1$\;
\While {$\mathit{now} < \mathit{terminate\text{-}at}$}
{
	$f_\text{old} \gets f(S)$\;
	$S \gets \mathcal{H}_h(S)$\;
    $f_\text{new} \gets f(S)$\;
    \If {$f_\text{new} > f_\text{old}$}
    {
	    $h \gets \mathit{RouletteWheel}(\Msucc_{h, 1}, \Msucc_{h, 2}, \ldots, \Msucc_{h, |\mathcal{H}|})$\;
        
        \If {$f(S) > f(S^*)$}
        {
        	$S^* \gets S$\;
        }
    }
	\Else
	{
	    $h \gets \mathit{RouletteWheel}(\Mfail_{h, 1}, \Mfail_{h, 2}, \ldots, \Mfail_{h, |\mathcal{H}|})$\;
    }
}

\Return {$S^*$}\;

\caption{Conditional Monte-Carlo Search}
\label{alg:cmchh}
\end{algorithm}

 CMCS does not in itself employ any learning during the search process, but is configured by means of offline learning, and so the behaviour of any specific instance of CMCS is defined by two matrices $\Msucc$ and $\Mfail$ of size $|\mathcal{H}| \times |\mathcal{H}|$ each.
 Thus, we refer to the general idea of CMCS as \emph{schema}, and to a concrete instance of CMCS, i.e.\ specific values of matrices $\Msucc$ and $\Mfail$, as \emph{configuration}.

 
 For the termination criterion, we use a predefined time after which CMCS terminates.
 This is most appropriate, as well as convenient, when we need to compare metaheuristics and in which different components run at different speeds so that simple counting of steps would not be a meaningful termination criterion. 

 CMCS requires an initial solution; this could have been supplied from one of the several construction heuristics developed for BBQP~\cite{Karapetyan2012,Duarte2014}, however, to reduce potential bias, we initialise the search with a randomly generated solution with probability of each of $x_i = 1$ and $y_j = 1$ being 50\%.

\subsection{CMCS properties}
 Below we list some of the properties of CMCS that make it a good choice in our study.
 We also believe that it will be useful in future studies in a similar way.
 \begin{itemize}
	\item 
    CMCS is able to combine several algorithmic components in one search process, and with each component taken as a black box.
	
    \item 
	CMCS has parameters for inclusion or exclusion of individual components as we do not know in advance if any of our components have poor performance.
	This is particularly true when considering that performance of a component might well depend on which others are available -- some synergistic combinations might be much more powerful than the individuals would suggest.
      
    \item 
    CMCS has parameters that permit some components to be used more often than others as some of our hill climbing operators are significantly faster than others; this also eliminates the necessity to decide in advance on the frequency of usage of each of the components. 
    Appropriate choices of the parameters should allow the imbalance of component runtimes to be exploited.
    
    \item 
    CMCS is capable of exploiting some (recent) history of the choices made by the metaheuristic, as there might be efficient sequences of components which should be exploitable.
    
    \item 
    As we will show later, CMCS is powerful enough to model some standard metaheuristics and, thus, allows easy comparison with standard approaches.
    
    \item
	The performance of CMCS does not depend on the absolute values of the objective function; it is rank-based in that it only uses the objective function to find out if a new solution is better than the previous solution.
	This property helps CMCS perform well across different families of instances.
	In contrast, methods such as Simulated Annealing, depend on the absolute values of the objective function and thus often need to be tuned for each family of instances, or else need some mechanism to account for changes to the scale of the objective function.
    
    \item 
    The transition matrices of a tuned CMCS configuration allow us conveniently interpret the results of automated generation.
\end{itemize}

\subsection{Special cases of CMCS}
\label{sec:special-cases}
 
 Several standard metaheuristics are special cases of CMCS\@.
 If $\mathcal{H} = \{ \text{HC}, \text{Mut} \}$ includes a hill climbing operator ``HC'' and a mutation ``Mut'' then
\begin{align*}
\Msucc = &\left(
	\begin{array}{ l | l l }
     				& \text{HC}	& \text{Mut} \\
     	\hline
        \text{HC}	& 1			& 0 \\
        \text{Mut}	& 1			& 0 \\
	\end{array}
\right)
\quad \text{and}\\
\Mfail = &\left(
	\begin{array}{ l | l l }
     				& \text{HC}	& \text{Mut} \\
     	\hline
        \text{HC}	& 0			& 1 \\
        \text{Mut}	& 1			& 0 \\
	\end{array}
\right)
\end{align*}
implements Iterated Local Search~\cite{Lourenco2010}; the algorithm repeatedly applies HC until it fails, then applies Mut, and then returns to HC disregarding the success or failure of Mut.
 
 If $\Msucc_{h,h'} = \Mfail_{h,h'} = 1 / |\mathcal{H}|$ for all $h, h' \in \mathcal{H}$ then CMCS implements a simple uniform random choice of component \cite{Cowling2001}.

 A generalisation of the uniform random choice is to allow non-uniform probabilities of component selection.
 We call this special case \emph{Operator Probabilities} (\opprob) and model it by setting $\Msucc_{h,h'} = \Mfail_{h,h'} = p_{h'}$ for some vector $p$ of probabilities.
 Note that Operator Probabilities is a static version of a Selection Hyper-heuristic~\cite{Cowling2001}.
 
 Obviously, if $\Msucc = \Mfail$ then CMCS implements a static version of MCHH\@.
 
 By allowing $\Msucc \neq \Mfail$, it is possible to implement a Variable Neighbourhood Search (VNS) using the CMCS schema.
 For example, if
 $$
 \Msucc = \left(
	 \begin{array}{ l | l l l l }
     				& \text{HC1}	& \text{HC2}	& \text{HC3}	& \text{\component{Mut}} \\
     	\hline
        \text{HC1}	& 1				& 0				& 0				& 0 \\
        \text{HC2}	& 1				& 0				& 0				& 0 \\
        \text{HC3}	& 1				& 0				& 0				& 0 \\
        \text{\component{Mut}}	& 1				& 0				& 0				& 0 \\
	 \end{array}
 \right)
 $$
 and
 $$
 \Mfail = \left(
	 \begin{array}{ l | l l l l }
     				& \text{HC1}	& \text{HC2}	& \text{HC3}	& \text{\component{Mut}} \\
     	\hline
        \text{HC1}	& 0				& 1				& 0				& 0 \\
        \text{HC2}	& 0				& 0				& 1				& 0 \\
        \text{HC3}	& 0				& 0				& 0				& 1 \\
        \text{\component{Mut}}	& 1				& 0				& 0				& 0 \\
	 \end{array}
 \right)
 $$
 then CMCS implements a VNS that applies HC1 until it fails, then applies HC2\@.
 If HC2 improves the solution then the search gets back to HC1; otherwise HC3 is executed.
 Similarly, if HC3 improves the solution then the search gets back to HC1; otherwise current solution is a local maximum with respect to the neighbourhoods explored by HC1, HC2 and HC3 (assuming they are deterministic) and mutation \component{Mut} is applied to diversify the search. 
 
 However, even though the previous examples are well-known metaheuristics, they are rather special cases from the perspective of CMCS, which allows much more sophisticated strategies. 
 For example, we can implement a two-loop heuristic, which alternates hill climbing operator HC1 and mutation Mut1 until HC1 fails to improve the solution.
 Then the control is passed to the second loop, alternating HC2 and Mut2.
 Again, if HC2 fails, the control is passed to the first loop.

 To describe such more sophisticated strategies, it is convenient to represent CMCS configurations with automata as in Figure~\ref{fig:two-loops}.
 Blue and red lines correspond to transitions in case of successful and unsuccessful execution of the components, respectively.
 Probabilities of each transition are shown with line widths (in Figure~\ref{fig:two-loops} all the shown probabilities are 100\%).
 The advantage of automata representation is that it visualises the probabilities of transition and sequences in which components are executed (and so complements, not supplants, the formal description via the pseudo-code and the explicit transition matrices), as common when describing transition systems.
 

\tikzset{vertex/.style={circle, draw, thick}}
\tikzset{edge base/.style={->, >=stealth'}}
\tikzset{improved/.style={edge base, blue, bend left=10}}
\tikzset{unimproved/.style={edge base, red, bend left=30}}
\tikzset{loop improved/.style={edge base, blue, loop above, in=70, out=110, looseness=8}}
\tikzset{loop unimproved/.style={edge base, red, loop above, in=60, out=120, looseness=9}}
 
\begin{figure}[htb]
	\centering
	\begin{tikzpicture}
		\node[vertex] (HC1) at (0, 0) {HC1};
		\node[vertex] (Mut1) at (0, 3) {\component{Mut1}};
		\node[vertex] (HC2) at (4, 0) {HC2};
		\node[vertex] (Mut2) at (4, 3) {\component{Mut2}};
		\path (HC1) edge[improved, ultra thick] (Mut1);
		\path (Mut1) edge[improved, ultra thick] (HC1);
		\path (Mut1) edge[unimproved, ultra thick] (HC1);
		\path (HC1) edge[unimproved, ultra thick, bend left=0] (Mut2);
		\path (HC2) edge[improved, ultra thick] (Mut2);
		\path (Mut2) edge[improved, ultra thick] (HC2);
		\path (Mut2) edge[unimproved, ultra thick] (HC2);
		\path (HC2) edge[unimproved, ultra thick, bend left=0] (Mut1);
	\end{tikzpicture}
	\caption{
    	Implementation of a two-loop heuristic within the CMCS framework.
        Blue lines show transitions in case of success, and red lines show transitions in case of failure of the component.
    }
	\label{fig:two-loops}
\end{figure}
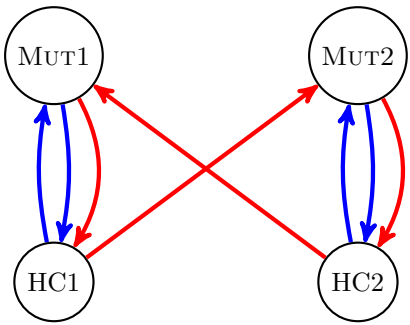

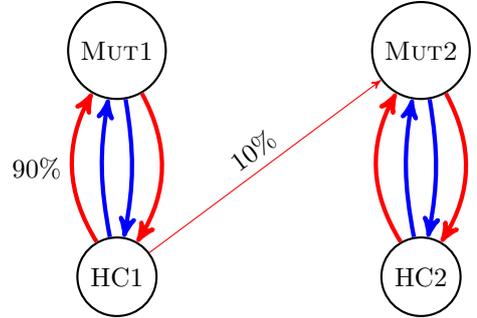
\begin{figure}[htb]
	\centering
	\begin{tikzpicture}
		\node[vertex] (HC1) at (0, 0) {HC1};
		\node[vertex] (Mut1) at (0, 3) {\component{Mut1}};
		\node[vertex] (HC2) at (4, 0) {HC2};
		\node[vertex] (Mut2) at (4, 3) {\component{Mut2}};
		\path (HC1) edge[improved, ultra thick] (Mut1);
		\path (Mut1) edge[improved, ultra thick] (HC1);
		\path (Mut1) edge[unimproved, ultra thick] (HC1);
		\path (HC1) edge[unimproved, ultra thick] node[midway, left, black] {90\%} (Mut1);
		\path (HC1) edge[unimproved, bend left=0] node[midway, above, sloped, black] {10\%} (Mut2);
		\path (HC2) edge[improved, ultra thick] (Mut2);
		\path (Mut2) edge[improved, ultra thick] (HC2);
		\path (Mut2) edge[unimproved, ultra thick] (HC2);
		\path (HC2) edge[unimproved, ultra thick] (Mut2);
	\end{tikzpicture}
	\caption{
    	Implementation of a two-phase heuristic with probabilistic transition from the first phase to the second phase.
        All the probabilities are 100\% unless otherwise labelled.
    }
	\label{fig:two-phase-probabilistic}
\end{figure}

 The transitions in the above example are deterministic, however, this is not an inherent limitation; for example, one could implement a two phase search with the transition being probabilistic, see Figure~\ref{fig:two-phase-probabilistic}.
 We also note here that CMCS can be significantly enriched by having several copies of each component in $\mathcal{H}$ and/or employing dummy components for describing more sophisticated behaviours; but we leave these possibilities to future work.

 These are just some of the options available with CMCS, showing that it is potentially a powerful tool.
 However, this flexibility does come with the associated challenge -- of configuring the matrices to generate effective metaheuristics.
 For example, if $|\mathcal{H}| = 10$ then CMCS has $2|\mathcal{H}|^2 = 200$ continuous parameters.
 
 By simple reasoning we can fix the values of a few of these parameters:
 \begin{itemize}
 	\item 
    If component $h$ is a deterministic hill climbing operator then $\Mfail_{h,h} = 0$, as when it fails then the solution remains unchanged and so immediate repetition is pointless.
    
    \item 
    If component $h$ is an idempotent operator (e.g.\ \OptX{} or \OptY{}) then $\Msucc_{h,h} = \Mfail_{h,h} = 0$; again there is no use in applying $h$ several times in a row.
 \end{itemize}
 
 Nevertheless, the significant number of remaining parameters of CMCS makes it hard to configure.
 For this reason we propose, and exploit a special case of the CMCS schema, with much fewer parameters but that still provides much of the power of the framework of the full CMCS\@.
 Specifically, we allow at most $k$ non-zero elements in each row of $\Msucc$ and $\Mfail$, calling the resulting metaheuristic ``\krow{$k$}''.
 Clearly, \krow{$|\mathcal{H}|$} is identical to the full version of CMCS\@.
 In practice, however, we expect one to use only smaller values of $k$; either $k = 1$ or $k = 2$.
  
 When $k = 1$, the corresponding automata has at most one outgoing ``success'' arc, and one outgoing ``failure'' arc for each component.
 Hence CMCS turns into a deterministic control mechanism.
 Note that iterated local search and VNS are in fact special cases of \krow{1}.
 
 When $k = 2$, the corresponding automata has at most two outgoing ``success'' arcs from each component, and their total probability of transition is 100\%.
 Hence, the ``success'' transition is defined by a pair of components and the split of probabilities between them.
 ``Failure'' transition is defined in the same way.
  
 In Section~\ref{sec:tuning}, we show that \krow{2} is sufficiently powerful to implement complex component combinations but is much easier to configure and analyse than full CMCS.
 
\tempclearpage
\section{Benchmark Instances}
\label{sec:testbed}

 The testbed which is currently de facto standard for BBQP was first introduced in our unpublished work~\cite{Karapetyan2012}.
 Our testbed consists of five instance types that correspond to some of the real life applications of BBQP\@.
 Here we provide the description of it, and also make it available for download.%
 \footnote{\url{http://csee.essex.ac.uk/staff/dkarap/?page=publications&key=CMCS-BBQP}}
 We keep record of the best known solutions for each of the test instances which will also be placed on the download page.

 In order to generate some of the instances, we need random bipartite graphs.
 To generate a random bipartite graph $G = (V, U, E)$, we define seven parameters, namely $m = |V|$, $n = |U|$, $\underline{d}_1$, $\bar{d_1}$, $\underline{d}_2$, $\bar{d_2}$ and $\mu$ such that $0 \le \underline{d}_1 \le \bar{d_1} \le n$, $0 \le \underline{d}_2 \le \bar{d_2} \le m$, $m \underline{d}_1 \le n \bar{d}_2$ and $m \bar{d}_1 \ge n \underline{d}_2$.

 The bipartite graph generator proceeds as follows.
\begin{enumerate}
	\item 
    For each node $v \in V$, select $d_v$ uniformly at random from the range $[\underline{d}_1, \bar{d}_1]$.
    
	\item 
    For each node $u \in U$, select $d_u$  uniformly at random from the range $[\underline{d}_2, \bar{d}_2]$.
	
    \item 
    While $\sum_{v \in V} d_v \neq \sum_{u \in U} d_u$, alternatively select a node in $V$ or $U$ and re-generate its degree as described above.%
    \footnote{In practice, if $m (\underline{d}_1 + \bar{d}_1) \approx n (\underline{d}_2 + \bar{d}_2)$, this algorithm converges very quickly.  
    However, in theory it may not terminate in finite time and, formally speaking, there needs to be a mechanism to guarantee convergence.
    Such a mechanism could be turned on after a certain (finite) number of unsuccessful attempts, and then it would force the changes of degrees $d_v$ that reduce $|\sum_{v \in V} d_v - \sum_{u \in U} d_u|$.
    }
    
	\item 
    Create a bipartite graph $G = (V, U, E)$, where $E = \emptyset$.
    
	\item 
    Randomly select a node $v \in V$ such that $d_v > \deg v$ (if no such node exists, go to the next step).  
    Let $U' = \{ u \in U \suchthat \deg u < d_u \connect{and} (v, u) \notin E \}$.  If $U' \neq \emptyset$, select a node $u \in U'$ randomly.  
    Otherwise randomly select a node $u \in U$ such that $(v, u) \notin E$ and $d_u > 0$; randomly select a node $v' \in V$ adjacent to $u$ and delete the edge $(v', u)$.  
    Add an edge $(v, u)$.  
    Repeat this step.
    
	\item 
    For each edge $(v, u) \in E$ select the weight $w_{vu}$ as a normally distributed integer with standard deviation $\sigma = 100$ and given mean $\mu$.
\end{enumerate}


The following are the instance types used in our computational experiments.

\begin{enumerate}
	\item 
    The \emph{Random} instances are as follows: $q_{ij}$, $c_i$ and $d_j$ are integers selected at random with normal distribution (mean $\mu = 0$ and standard deviation $\sigma = 100$).

	\item 
    The \emph{Max Biclique} instances model the problem of finding a biclique of maximum weight in a bipartite graph.  
    Let $G = (I, J, E)$ be a random bipartite graph with $\underline{d}_1 = n / 5$, $\bar{d_1} = n$, $\underline{d}_2 = m / 5$, $\bar{d_2} = m$ and $\mu = 100$. 
    (Note that setting $\mu$ to 0 would make the weight of any large biclique likely to be around 0, which would make the problem much easier.)
    If $w_{ij}$ is the weight of an edge $(i, j) \in E$, set $q_{ij} = w_{ij}$ for every $i \in I$ and $j \in J$ if $(i, j) \in E$ and $q_{ij} = -M$ otherwise, where $M$ is large number.  
    Set $c$ and $d$ as zero vectors.

	\item 
    The \emph{Max Induced Subgraph} instances model the problem of finding a subset of nodes in a bipartite graph that maximises the total weight of the induced subgraph.  
    The Max Induced Subgraph instances are similar to the Max Biclique instances except that $q_{ij} = 0$ if $(i, j) \notin E$ and $\mu = 0$.  
    (Note that if $\mu > 0$ then the optimal solution would likely include all or almost all the nodes and, thus, the problem would be relatively easy).

	\item 
    The \emph{MaxCut} instances model the MaxCut problem as follows.  
    First, we generate a random bipartite graph as for the Max Induced Subgraph instances.  
    Then, we set $q_{ij} = -2 w_{ij}$ if $(i, j) \in E$ and $q_{ij} = 0$ if $(i, j) \notin E$.  
    Finally, we set $c_i = \frac{1}{2} \sum_{j \in J} q_{ij}$ and $d_j = \frac{1}{2} \sum_{i \in I} q_{ij}$.  
    For an explanation, see~\cite{Punnen2012}.

	\item The \emph{Matrix Factorisation} instances model the problem of producing a rank one approximation of a binary matrix.  The original matrix $H = (h_{ij})$ (see Section~\ref{sec:introduction}) is generated randomly with probability 0.5 of $h_{ij} = 1$.  The values of $q_{ij}$ are then calculated as $q_{ij} = 1 - 2 h_{ij}$, and $c$ and $d$ are zero vectors.
\end{enumerate}

 Our benchmark consists of two sets of instances: Medium and Large.
 Each of the sets includes one instance of each type (Random, Max Biclique, Max Induced Subgraph, MaxCut and Matrix Factorisation) of each of the following sizes:\\
 Medium: $200 \times 1000$, $400 \times 1000$, $600 \times 1000$, $800 \times 1000$, $1000 \times 1000$;\\
 Large: $1000 \times 5000$, $2000 \times 5000$, $3000 \times 5000$, $4000 \times 5000$, $5000 \times 5000$.\\
 Thus, in total, the benchmark includes 25 medium and 25 large instances.

\tempclearpage
\section{Metaheuristic Design}
\label{sec:tuning}

 In this section we describe configuration of metaheuristics as discussed in Section~\ref{sec:cmchh} and using the BBQP components given in Section~\ref{sec:components}.
 In Sections~\ref{sec:solution-representation} and~\ref{sec:polishing} we give some details about our experiments, then in Section~\ref{sec:metaheuristic-generation} describe the employed automated configuration technique, in Section~\ref{sec:emergent-metaheuristics} we provide details of the configured metaheuristics, and in Section~\ref{sec:analysis} analyse the results.
 
 Our test machine is based on two Intel Xeon CPU E5-2630 v2 (2.6~GHz) and has 32~GB RAM installed.  
 Hyper-threading is enabled, but we never run more than one experiment per physical CPU core concurrently, and concurrency is not exploited in any of the tested solution methods. 
 
\subsection{Solution Representation}
\label{sec:solution-representation}

We use the most natural solution representation for BBQP, i.e. simply storing vectors $x$ and $y$.
 However,  additionally storing some auxiliary information with the solution can dramatically improve the performance of algorithms.
 We use a strategy similar to the one employed in~\cite{Glover2015}.
 In particular, along with vectors $x$ and $y$, we always maintain values $c_i + \sum_j y_j q_{ij}$ for each $i$, and $d_j + \sum_i x_i q_{ij}$ for each $j$.
 Maintenance of this auxiliary information slows down any updates of the solution but significantly speeds up the evaluation of potential moves, which is what usually takes most of time during the search.
 
\subsection{Solution Polishing}
\label{sec:polishing}

 As in many single-point metaheuristics, the changes between diversifying and intensifying steps of CMCS mean that the best found solution needs to be stored, and also that it is not necessarily a local maximum with respect to all the available hill climbing operators.
 Hence, we apply a polishing procedure to every CMCS configuration produced in this study, including special cases of VNS, \opprob{}\ and MCHH\@.
 Our polishing procedure is executed after the CMCS finishes its work, and it is aimed at improving the best solution found during the run of CMCS\@.
 It sequentially executes \OptX{}, \OptY{}, \FlpX{} and \FlpY{} components, restarting this sequence every time an improvement is found.
 When none of these algorithms can improve the solution, that is, the solution is a local maximum with respect to all of our hillclimbing operators, the procedure terminates.

 While taking very little time, this polishing procedure has notably improved our results.
 We note that this polishing stage is a Variable Neighbourhood Descent, and thus a special case of CMCS; hence, the final polishing could be represented as a second phase of CMCS\@.
 We also note that the Tabu Search algorithm, against which we compare our best CMCS configuration in Section~\ref{sec:comparison}, uses an equivalent polishing procedure applied to each solution and thus the comparison is fair.

\subsection{Approach to Configuration of the Metaheuristics }
\label{sec:metaheuristic-generation}

 Our ultimate goal in this experiment is to apply automated configuration (e.g.\ in the case of CMCS, to configure $\Msucc$ and $\Mfail$ matrices), which would compete with the state-of-the-art methods on the benchmark instances (which have sizes $200 \times 1000$ to $5000 \times 5000$) and with running times in the order of several seconds to several minutes.
 As explained in Section~\ref{sec:cmchh}, instead of hand designing a metaheuristic we chose to use automated generation based on the CMCS schema.
 Automated generation required a set of training instances.
 Although straightforward, directly training on benchmark instances would result in over-training (a practice generally considered unfair because an over-trained heuristic might perform well only on a very small set of instances on which it is tuned and then tested) and also would take considerable computational effort.
 Thus, for training we use instances of size $200 \times 500$.
 We also reduced the running times to 100~milliseconds per run of each candidate configuration, that is, matrices when configuring CMCS or MCHH, probability vector for \opprob{}, and component sequence for VNS.
 
 Let $T$ be the set of instances used for training.
 Then our objective function for configuration is
\begin{equation}
 \label{eq:tuning-objective}
 f(h, T) = \frac{1}{|T|} \sum_{t \in T} \frac{f_\text{best}(t) - h(t)}{f_\text{best}(t)} \cdot 100\% \,,
\end{equation}
where $h$ is the evaluated heuristic, $h(t)$ is the objective value of solution obtained by $h$ for instance $t$, and $f_\text{best}(t)$ is the best known solution for instance $t$.
 For the training set, we used instances of all of the types.
 In particular, we use one instance of each of the five types (see Section~\ref{sec:testbed}), all of size $200 \times 500$, and each of these training instances is included in $T$ 10 times, thus $|T| = 50$ (we observed that without including each instance several times the noise level significantly obfuscated results).
 Further, when testing the top ten candidates, we include each of the five instances 100 times in $T$, thus having $|T| = 500$. 
 
 We consider four types of metaheuristics: VNS, \opprob,\ MCHH and \krow{2}, all of which are also special cases of CMCS\@.
 All the components discussed in Section~\ref{sec:components}, and also briefly described in Table~\ref{tab:components}, are considered for inclusion in all the metaheuristics.
 Additionally, since \Repair{} is a totally new component, we want to confirm its usefulness.
 For this we also study a special case of \krow{2} which we call ``\emph{\reduced}''.
 In \reduced{}, the pool of potential components includes all the components in Table~\ref{tab:components} except \Repair{}. 

To configure VNS and \opprob{}, we use brute force search as we can reasonably restrict the search to a relatively small number of options.
 In particular, when configuring \opprob{}, the number of components $|\mathcal{H}|$ (recall that $\mathcal{H}$ is the set of components employed by the metaheuristic) is restricted to at most four, and weights of individual components are selected from $\{ 0.1 \commanospace 0.2 \commanospace 0.5 \commanospace 0.8 \commanospace 1 \}$ (these weights are then rescaled to obtain probabilities).
 We also require that there has to be at least one hill climbing operator in $\mathcal{H}$ as otherwise there would be no pressure to improve the solution, and one mutation operator as otherwise the search would quickly become trapped in a local maximum.
 Note that we count \Repair{} as a mutation as, although designed to explicitly fix flaws, it is quite likely to worsen the solution (even if in the long run this will be beneficial).
 When configuring VNS, $\mathcal{H}$ includes one or several hill climbing operators and one mutation and the configuration process has to also select the order in which they are applied.

 To configure CMCS and static MCHH, we use a simple evolutionary algorithm, with the solution describing matrices $\Msucc$ and $\Mfail$ (accordingly restricted), and fitness function (\ref{eq:tuning-objective}).
 Implementation of a specialised tuning algorithm has an advantage over the general-purpose automated algorithm configuration packages, as a specialised system can exploit the knowledge of the parameter space (such as entanglement of certain parameters).
 In this project, our evolutionary algorithm employs specific neighbourhood operators that intuitively make sense for this particular application. 
 For example, when tuning 2-row, we employ, among others, a mutation operator that swaps the two non-zero weights in a row of a weight matrix.
 Such a move is likely to be useful for ``exploitation''; however it is unlikely to be discovered by a general purpose parameter tuning algorithm.
 
 We compared the tuning results of our CMCS-specific algorithm to ParamILS~\cite{HutterEtal2009:ParamILS}, one of the leading general purpose automated parameter tuning/algorithm configuration software.
 We found out that, while ParamILS performs well, our specialised algorithm clearly outperforms it, producing much better configurations.
 It should be noted that there can be multiple approaches to encode matrices $\Msucc$ and $\Mfail$ for ParamILS.
 We tried two most natural approaches and both attempts were relatively unsuccessful; however it is possible that future research will reveal more efficient ways to represent the key parameters of CMCS\@.
 We also point out that CMCS can be a new interesting benchmark for algorithm configuration or parameter tuning software.

\subsection{Configured Metaheuristics}
\label{sec:emergent-metaheuristics}

 In this section we describe the configurations of each type (VNS, \opprob{}, MCHH, \reduced{} and \krow{2}) generated as described in Section~\ref{sec:metaheuristic-generation}.
 From now on we refer to the obtained configurations by the name of their types.
 Note that the structures described in this section are all machine-generated, and thus when we say that ``a metaheuristic chose to do something'', we mean that such a decision emerged from the generation process; the decision was not a human choice.

\bigskip
 
VNS chose three hill climbing operators, \OptY{}, \FlpY{} and \OptX{}, and a mutation \MutX{16}, and using the order as written.
 It is interesting to observe that this choice and sequence can be easily explained.
 Effectively, the search optimises $y$ given a fixed $x$ (\OptY{}), then tries small changes to $x$ with some lookahead (\FlpY{}), and if this fails then optimises $x$ globally but without lookahead (\OptX{}).
 If the search is in a local maximum with respect to all three neighbourhoods then the solution is perturbed by a strong mutation \MutX{16}.
 Observe that the sequence of hill climbing operators does not obey the generally accepted rule of thumb to place smaller neighbourhoods first; the third hill climbing operator \OptX{} has clearly smaller neighbourhood than \FlpY{}\@.
 However, this sequence has an interesting internal logic.
 Whenever \FlpY{} succeeds in improving the solution, the resultant solution is a local minimum with respect to \OptX{}.
 Accordingly, VNS jumps back to \OptY{} when \FlpY{} succeeds.
 However, if \FlpY{} fails then the solution might not be a local minimum with respect to \OptX{}, and then \OptX{} is executed.
 This shows that the automated configuration is capable of generating meaningful configurations which are relatively easy to explain but might not be so easy to come up with.

\bigskip

 The \opprob{}\ chose four components: \OptX{} (probability of picking is 40\%), \FlpX{} (20\%), \Repair{} (20\%) and \MutX{16} (20\%).
 Note that the actual runtime frequency of \OptX{} is only about 30\% because the framework will never execute \OptX{} twice in a row.

\bigskip

 Out of 9 components, MCHH chose five: \OptX{}, \OptY{}, \FlpX{}, \MutY{4} and \MutX{16}.
 The generated transition matrix (showing the probabilities of transitions) is given in Figure~\ref{fig:mchh-matrix}.
 
\begin{figure}[htb]
 $$
 \begin{array}{@{} r | rrrrr @{}}
	& \text{\OptX{}} & \text{\OptY{}} & \text{\FlpX{}} & \text{\MutY{4}} & \text{\MutX{16}} \\
    \midrule
	\text{\OptX{}} &-	&78.2\%	&5.4\%	&12.9\%	&3.5\% \\
	\text{\OptY{}} &86.9\%	&-	&0.0\%	&13.1\%	&0.0\% \\
	\text{\FlpX{}} &16.2\%	&30.1\%	&19.4\%	&5.0\%	&29.3\% \\
	\text{\MutY{4}} &35.6\%	&24.7\%	&22.9\%	&4.1\%	&12.8\% \\
	\text{\MutX{16}} &1.4\%	&84.6\%	&0.0\%	&14.0\%	&0.0\% \\
 \end{array}
 $$
 
\caption{
	Transition matrix of MCHH\@.
    Dashes show prohibited transitions, i.e.\ the transitions that are guaranteed to be useless and so are constrained to zero, as opposed to being set to zero by the tuning generation process.
    In this table, and subsequent ones, the row specifies the previously executed component, and the column specifies the next executed component.
}
\label{fig:mchh-matrix}
\end{figure}

\bigskip

 \reduced{} chose to use only \OptX{}, \OptY{}, \FlpX{}, \MutX{4}, \MutY{4} and \MutY{16} from the pool of 8 components it was initially permitted (recall that \reduced{} was not allowed to use \Repair{}), and transition matrices as given in Figure~\ref{fig:reduced-matrices} and visually illustrated in Figure~\ref{fig:reduced}.
 The line width in Figure~\ref{fig:reduced} indicates the frequency of the transition when we tested the configuration on the tuning instance set.
 Although these frequencies may slightly vary depending on the particular instance, showing frequencies preserves all the advantages of showing probabilities but additionally allows one to see: (i) how often a component is executed (defined by the total width of all incoming/outgoing arrows), (ii) the probability of success of a component (defined by the the total width of blue outgoing arrows compared to the total width of the red outgoing arrows), and (iii) most common sequences of component executions (defined by thickest arrows).
 
\begin{figure*}[htb]
\small
\begin{subfigure}{0.48\textwidth}
 $$
 \begin{array}{@{} r | *{6}{@{~~}r} @{}}
 	& \text{\OptX{}} & \text{\OptY{}} & \text{\FlpX{}} & \text{\MutX{4}} & \text{\MutY{4}} & \text{\MutY{16}} \\
    \midrule
	\text{\OptX{}}	&-	&100\%	&.	&.	&.	&.	\\
	\text{\OptY{}}	&8\%	&-	&.	&.	&92\%	&.	\\
	\text{\FlpX{}}	&38\%	&62\%	&.	&.	&.	&.	\\
	\text{\MutX{4}}	&.	&.	&.	&100\%	&.	&.	\\
	\text{\MutY{4}}	&45\%	&.	&.	&.	&.	&55\%	\\
	\text{\MutY{16}}	&.	&.	&.	&54\%	&.	&46\%	\\
 \end{array}
 $$
 \caption{$\Msucc$}
\end{subfigure}
\hfill
\begin{subfigure}{0.48\textwidth}
 $$
 \begin{array}{@{} r | *{6}{@{~~}r} @{}}
 	& \text{\OptX{}} & \text{\OptY{}} & \text{\FlpX{}} & \text{\MutX{4}} & \text{\MutY{4}} & 	\text{\MutY{16}} \\
    \midrule
	\text{\OptX{}}	&-	&.	&.	&.	&100\%	&.	\\
	\text{\OptY{}}	&.	&-	&80\%	&20\%	&.	&.	\\
	\text{\FlpX{}}	&.	&.	&-	&.	&.	&100\%	\\
	\text{\MutX{4}}	&28\%	&.	&72\%	&.	&.	&.	\\
	\text{\MutY{4}}	&68\%	&.	&.	&.	&.	&32\%	\\
	\text{\MutY{16}}	&51\%	&.	&.	&.	&.	&49\%	\\
 \end{array}
 $$
 \caption{$\Mfail$}
\end{subfigure}

\caption{
	Transition matrices of \reduced{}.
	Dashes show prohibited transitions, see Section~\ref{sec:special-cases}. 
    \reduced{} transition frequencies are shown in Figure~\ref{fig:reduced}.
}
\label{fig:reduced-matrices}
\end{figure*}

\bigskip

 \krow{2} decided to use only  \OptX{}, \OptY{}, \FlpX{}, \Repair{}, \MutY{4} and \MutY{16} from the set of 9 moves it was initially permitted, and transition matrices as shown in Figure~\ref{fig:tworow-matrices}.
 
\begin{figure*}[htb]
\small
\begin{subfigure}{0.48\textwidth}
 $$
 \begin{array}{@{} r | *{6}{@{~~}r} @{}}
 	& \text{\OptX{}} & \text{\OptY{}} & \text{\FlpX{}} & \text{\Repair{}} & \text{\MutY{4}} & 	\text{\MutY{16}} \\
    \midrule    
	\text{\OptX{}}	&-	&66\%	&.	&.	&.	&34\%	\\
	\text{\OptY{}}	&41\%	&-	&.	&.	&59\%	&.	\\
	\text{\FlpX{}}	&.	&29\%	&71\%	&.	&.	&.	\\
	\text{\Repair{}}	&41\%	&.	&.	&.	&.	&59\%	\\
	\text{\MutY{4}}	&.	&.	&40\%	&60\%	&.	&.	\\
	\text{\MutY{16}}	&.	&.	&.	&55\%	&45\%	&.	\\
 \end{array}
 $$
 \caption{$\Msucc$}
\end{subfigure}
\hfill
\begin{subfigure}{0.48\textwidth}
 $$
 \begin{array}{@{} r | *{6}{@{~~}r} @{}}
 	& \text{\OptX{}} & \text{\OptY{}} & \text{\FlpX{}} & \text{\Repair{}} & \text{\MutY{4}} & 	\text{\MutY{16}} \\
    \midrule
	\text{\OptX{}}	&-	&.	&.	&.	&100\%	&.	\\
	\text{\OptY{}}	&25\%	&-	&.	&75\%	&.	&.	\\
	\text{\FlpX{}}	&45\%	&55\%	&-	&.	&.	&.	\\
	\text{\Repair{}}	&2\%	&.	&.	&.	&.	&98\%	\\
	\text{\MutY{4}}	&87\%	&.	&.	&13\%	&.	&.	\\
	\text{\MutY{16}}	&.	&.	&.	&.	&82\%	&18\%	\\
 \end{array}
 $$
 \caption{$\Mfail$}
\end{subfigure}

\caption{
	Transition matrices of \krow{2}, our best performing metaheuristic.
	Dashes show prohibited transitions.
   \krow{2} transition frequencies are shown in Figure~\ref{fig:tworow}.
}
\label{fig:tworow-matrices}
\end{figure*}

\subsection{Analysis of Components and Metaheuristics}
\label{sec:analysis}

\begin{table}[htb]
\centering
\small
\begin{tabular}{@{} l r r @{}}
\toprule
Metaheuristic 
	& Objective value (\ref{eq:tuning-objective})
    & Comp.\ exec.
    \\
    
\midrule
    
VNS
	& 0.598\%
    & 384
    \\    
    

    

\opprob{}
	& 0.448\%
    & 520
    \\


MCHH
	& 0.395\%
    & 2008
    \\    
    
\reduced{}
	& 0.256\%
    & 5259
    \\
    
\krow{2}
	& 0.242\%
    & 5157
    \\
    
\bottomrule
\end{tabular}
\caption{
	Performance of the emergent metaheuristics on the training instance set.
    Rows are ordered by performance of metaheuristics, from worst to best.
}
\label{tab:heuristics500}
\end{table}

 Table~\ref{tab:heuristics500} gives the tuning objective function (\ref{eq:tuning-objective}) and the average number of component executions per run (i.e.\ in 100 milliseconds when solving a $200 \times 500$ instance) for each metaheuristic.
 CMCS, even if restricted to \krow{2} and even if the pool of components is reduced, outperforms all standard metaheuristics (VNS, \opprob{}\ and MCHH), even though \opprob{}\ and VNS benefit from higher quality configuration (recall that VNS and \opprob{} are configured using complete brute-force search).
 An interesting observation is that the best performing metaheuristics mostly employ fast components thus being able to run many more iterations than, say, VNS or \opprob{}
 
\begin{figure*}[htb]
\includegraphics[scale=1]{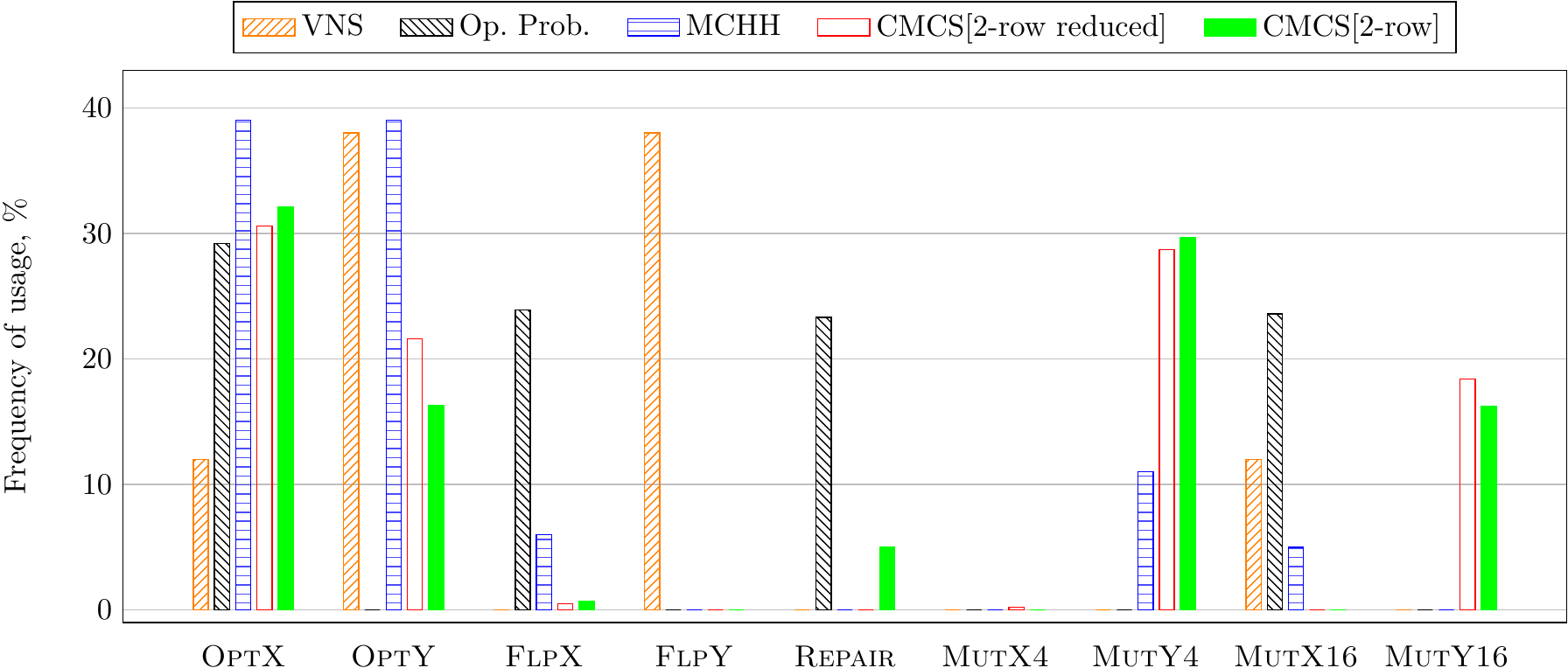}

\caption{
	Runtime frequency of usage of the components in tuned metaheuristics.
}
\label{fig:components}
\end{figure*}

 Figure~\ref{fig:components} gives the relative frequency of usage of each component by each metaheuristic.
 Most of the components appear to be useful within at least one of the considered metaheuristic schemas; only \MutX{4} is almost unused.
 It is however not surprising to observe some imbalance between the Mutation-X and Mutation-Y components because the number of rows is about half of the number of columns in the training instances.
 The selection of components is hard to predict as it significantly depends on the metaheuristic schema; indeed, different types of metaheuristics may be able to efficiently exploit different features of the components.
 Thus components should not be permanently discarded or selected based only on expert intuition and/or a limited number of experiments.
 We believe that the approach to component usage analysis proposed and used in this paper (and also in works such as \cite[and others]{HutterEtal2009:ParamILS,BezerraEtal2015:component-MOO}) is in many circumstances more comprehensive than manual analysis.

 While frequencies of usage of the components vary between all the metaheuristics, \opprob{}\ is clearly an outlier in this respect.  
 We believe that this reflects the fact that \opprob{}\ is the only metaheuristic among the considered ones that does not have any form of memory and thus does not control the order of components.
 Thus it prefers strong (possibly slow) components whereas other metaheuristics have some tendency to form composite components from fast ones, with the latter (history-based) approach apparently being superior.

\begin{figure*}[htb]
\begin{subfigure}{0.45\textwidth}
\centering
\begin{tikzpicture}[
	scale=0.7
    ]
	\node[vertex] (OptX) at   ( 4,  0) {\OptX{}};
	\node[vertex] (OptY) at   ( 7, -4) {\OptY{}};
	\node[vertex] (FlpX) at   ( 1, -2) {\FlpX{}};
	\node[vertex] (MutX4) at  ( 1,  2) {\MutX{4}};
	\node[vertex] (MutY4) at  (10,  0) {\MutY{4}};
	\node[vertex] (MutY16) at ( 7,  4) {\MutY{16}};
	\path (OptX) edge[improved=100, line width=5, bend right=10] (OptY);
	\path (OptX) edge[unimproved=100, line width=3.28167042045076, bend right=10] (MutY4);
	\path (OptY) edge[improved=100, line width=1.42334020373296, bend right=10] (OptX);
	\path (OptY) edge[unimproved=100, line width=0.651997311703321, bend left=5] (FlpX);
	\path (OptY) edge[unimproved=100, line width=0.306307821512429, bend left=15] (MutX4);
	\path (OptY) edge[improved=100, line width=4.76918229710326, bend right=10] (MutY4);
	\path (FlpX) edge[improved=100, line width=0.410522026332434, bend right=10] (OptX);
	\path (FlpX) edge[improved=100, line width=0.542967234955163, bend right=20] (OptY);
	\path (FlpX) edge[unimproved=100, line width=0.353484210062046, bend left=25, in=140] (MutY16);
	\path (MutX4) edge[unimproved=100, line width=0.270868704757155, bend left=10] (OptX);
	\path (MutX4) edge[unimproved=100, line width=0.403970722495793, bend right=10] (FlpX);
	\path (MutY4) edge[improved=100, line width=0.552304926092725, bend right=30] (OptX);
	\path (MutY4) edge[unimproved=100, line width=4.7281679910585, bend right=10] (OptX);
	\path (MutY4) edge[improved=100, line width=0.62090678787576, bend right=10] (MutY16);
	\path (MutY4) edge[unimproved=100, line width=3.23389775596361, bend right=30] (MutY16);
	\path (MutY16) edge[unimproved=100, line width=3.28979036462274, bend right=30] (OptX);
	\path (MutY16) edge[improved=100, line width=0.377806456179487, bend right=30] (MutX4);
	\path (MutY16) edge[loop improved=100, line width=0.35473993057108] (MutY16);
	\path (MutY16) edge[loop unimproved=100, line width=3.23300387268048] (MutY16);
\end{tikzpicture}
\caption{
	Emergent \reduced{}, i.e.\ a metaheuristic which was allowed to use any components except \Repair{}.
}
\label{fig:reduced}
\end{subfigure}
\hfill
\begin{subfigure}{0.50\textwidth}
\centering
\begin{tikzpicture}[
	scale=0.7
    ]
	\node[vertex] (OptX) at   ( 4,  0) {\OptX{}};
	\node[vertex] (OptY) at   ( 7, -4) {\OptY{}};
	\node[vertex] (FlpX) at   ( 2, -2) {\FlpX{}};
	\node[vertex] (MutY4) at  (10,  0) {\MutY{4}};
	\node[vertex] (MutY16) at ( 7,  4) {\MutY{16}};
	\node[vertex] (Repair) at (13,  3) {\Repair{}};
	\path (OptX) edge[improved=100, line width=3.98816860100795, bend right=10] (OptY);
	\path (OptX) edge[unimproved=100, line width=2.72247442389867, bend right=10] (MutY4);
	\path (OptX) edge[improved=100, line width=2.91259937316483, bend left=10] (MutY16);
	\path (OptY) edge[improved=100, line width=2.53561881907278, bend right=10] (OptX);
	\path (OptY) edge[unimproved=100, line width=0.366936388326369, bend right=30] (OptX);
	\path (OptY) edge[unimproved=100, line width=0.632518787075294, bend right=40] (Repair);
	\path (OptY) edge[improved=100, line width=3.03028146414151, bend right=10] (MutY4);
	\path (FlpX) edge[unimproved=100, line width=0.191169481593167, bend left=10] (OptX);
	\path (FlpX) edge[improved=100, line width=0.434607789428906, bend right=15] (OptY);
	\path (FlpX) edge[unimproved=100, line width=0.230407336222265, bend right=30] (OptY);
	\path (FlpX) edge[loop improved=100, line width=0.674033851937335] (FlpX);
	\path (Repair) edge[improved=100, line width=0.322880193809841, bend right=15] (OptX);
	\path (Repair) edge[unimproved=100, line width=0.282871000989214, bend right=30] (OptX);
	\path (Repair) edge[improved=100, line width=0.396660005722433, bend right=15] (MutY16);
	\path (Repair) edge[unimproved=100, line width=2.15320966878754, bend right=30] (MutY16);
	\path (MutY4) edge[unimproved=100, line width=5, bend right=10] (OptX);
	\path (MutY4) edge[improved=100, line width=0.531017394082002, bend left=30] (FlpX);
	\path (MutY4) edge[improved=100, line width=0.643373095205325, bend right=10] (Repair);
	\path (MutY4) edge[unimproved=100, line width=1.9435261563258, bend right=30] (Repair);
	\path (MutY16) edge[improved=100, line width=0.623020508480837, bend left=45] (Repair);
	\path (MutY16) edge[improved=100, line width=0.535346411863011, bend left=10] (MutY4);
	\path (MutY16) edge[unimproved=100, line width=3.54967778117428, bend left=25] (MutY4);
	\path (MutY16) edge[loop unimproved=100, line width=1.67072663720565] (MutY16);
\end{tikzpicture}
\caption{
	Emergent \krow{2}, i.e.\ our best performing metaheuristic which was allowed to use any components.
}
\label{fig:tworow}
\end{subfigure}

\caption{
	Runtime frequencies of \reduced{} and \krow{2} tested on the training instance set.
	The names and brief descriptions of each component are given in Table~\ref{tab:components}.
}
\label{fig:cmcs-tuned}
\end{figure*}
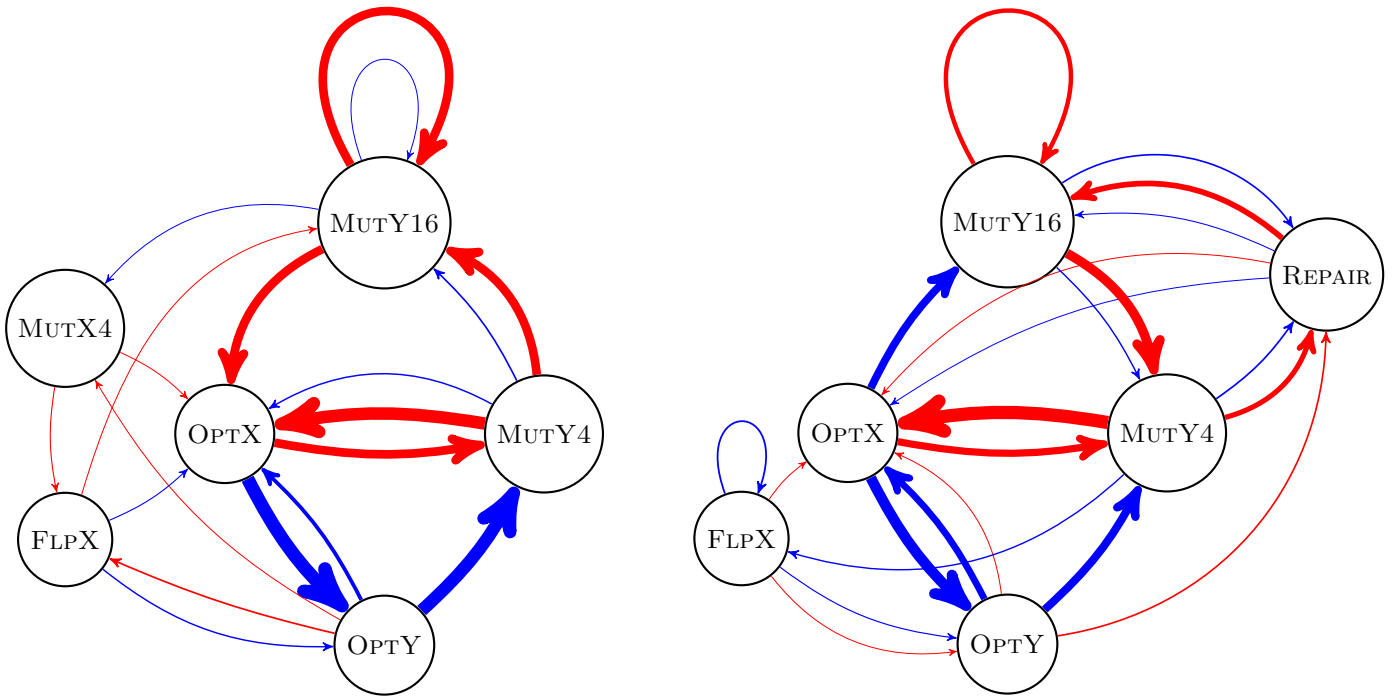

 More information about the performance of \reduced{} and \krow{2} configurations can be collected from Figure~\ref{fig:cmcs-tuned} detailing the runtime frequencies of transitions in each of them.
 Edge width here is proportional to square root of the runtime frequency of the corresponding transition occurring in several test runs; thus it allows to see not only the probabilities of transitions from any individual component, but also how frequently that component was executed and how often it was successful, compared to other components. 
 
 Firstly, we observe that the two metaheuristics employ similar sets of components; the only difference is that \krow{2} does not use \MutX{4} but adds \Repair{} (recall that \Repair{} was purposely removed from the pool of components of \reduced{}).
 Furthermore, the core components (\OptX{}, \OptY{}, \MutY{4} and \MutY{16}) are exactly the same, and most of interconnections between them are similar.
 However, the direction of transitions to and from \MutY{16} is different.
 One may also notice that both metaheuristics have ``mutation'' blocks; that is, mutations that are often executed in sequences.
 It is then not surprising that \krow{2} connects \Repair{} to the other mutation components.

 Both metaheuristics include some natural patterns such as alternation of \OptX{} and \OptY{}, or iterated local search \OptX{}--\MutY{4}, which we could also expect in a hand-designed metaheuristic.
 It is also easy to suggest an explanation for the loop at \MutY{16} as it allows the component to be repeated a couple of times intensifying the mutation.
 However, the overall structure of the metaheuristics is complex and hard to explain.
 Our point here is that, although the observed chains of components make sense, it is unlikely that a human expert would come up with a heuristic of such a level of detail.

\tempclearpage
\section{Evaluation of Metaheuristics}
\label{sec:evaluation}

\begin{table*}[htb]
\centering
\begin{tabular}{@{} lrrrrr @{}}
\toprule
Instance & VNS & \opprob{} & MCHH & \reduced{} & \krow{2} \\

\midrule																	Rand 200x1000			&		{{0.01}}		&		\underline{\textbf{0.00}}		&		\underline{\textbf{0.00}}		&		\underline{\textbf{0.00}}		&		\underline{\textbf{0.00}}		\\
Rand 400x1000			&		{{0.05}}		&		\underline{\textbf{0.00}}		&		\underline{\textbf{0.00}}		&		\underline{\textbf{0.00}}		&		\underline{\textbf{0.00}}		\\
Rand 600x1000			&		\underline{\textbf{0.00}}		&		\underline{\textbf{0.00}}		&		{{0.00}}		&		\underline{\textbf{0.00}}		&		\underline{\textbf{0.00}}		\\
Rand 800x1000			&		{{0.08}}		&		\underline{\textbf{0.00}}		&		{{0.01}}		&		{{0.00}}		&		{{0.00}}		\\
Rand 1000x1000			&		{{0.07}}		&		{{0.03}}		&		{{0.20}}		&		\underline{\textbf{0.00}}		&		{{0.04}}		\\[1ex]
																							
Biclique 200x1000			&		{{0.88}}		&		\underline{\textbf{0.00}}		&		\underline{\textbf{0.00}}		&		\underline{\textbf{0.00}}		&		\underline{\textbf{0.00}}		\\
Biclique 400x1000			&		\underline{\textbf{0.00}}		&		\underline{\textbf{0.00}}		&		{{0.14}}		&		{{0.09}}		&		{{0.09}}		\\
Biclique 600x1000			&		{\textbf{0.09}}		&		{{0.54}}		&		{{0.95}}		&		{{0.55}}		&		{{1.48}}		\\
Biclique 800x1000			&		\underline{\textbf{0.00}}		&		{{0.53}}		&		{{0.34}}		&		{{0.24}}		&		{{0.56}}		\\
Biclique 1000x1000			&		\underline{\textbf{0.00}}		&		{{0.14}}		&		{{0.13}}		&		{{0.16}}		&		{{0.35}}		\\[1ex]
																							
MaxInduced 200x1000			&		\underline{\textbf{0.00}}		&		\underline{\textbf{0.00}}		&		\underline{\textbf{0.00}}		&		\underline{\textbf{0.00}}		&		\underline{\textbf{0.00}}		\\
MaxInduced 400x1000			&		\underline{\textbf{0.00}}		&		\underline{\textbf{0.00}}		&		\underline{\textbf{0.00}}		&		\underline{\textbf{0.00}}		&		\underline{\textbf{0.00}}		\\
MaxInduced 600x1000			&		{{0.18}}		&		\underline{\textbf{0.00}}		&		\underline{\textbf{0.00}}		&		\underline{\textbf{0.00}}		&		\underline{\textbf{0.00}}		\\
MaxInduced 800x1000			&		{{0.30}}		&		{{0.08}}		&		{{0.08}}		&		{{0.09}}		&		{\textbf{0.00}}		\\
MaxInduced 1000x1000			&		{{0.16}}		&		{{0.04}}		&		{{0.04}}		&		{{0.04}}		&		{\textbf{0.03}}		\\[1ex]
																							
BMaxCut 200x1000			&		{{1.76}}		&		{{0.14}}		&		{{0.09}}		&		{{0.43}}		&		{\textbf{0.06}}		\\
BMaxCut 400x1000			&		{{2.25}}		&		{{0.67}}		&		{{1.25}}		&		{{0.89}}		&		{\textbf{0.40}}		\\
BMaxCut 600x1000			&		{{2.46}}		&		{{1.18}}		&		{{3.19}}		&		{{1.16}}		&		{\textbf{0.53}}		\\
BMaxCut 800x1000			&		{{4.35}}		&		{{2.19}}		&		{{2.75}}		&		{{1.49}}		&		{\textbf{1.05}}		\\
BMaxCut 1000x1000			&		{{4.51}}		&		{{2.65}}		&		{{2.39}}		&		{\textbf{0.39}}		&		{{0.46}}		\\[1ex]
																							
MatrixFactor 200x1000			&		\underline{\textbf{0.00}}		&		{{0.27}}		&		{{0.05}}		&		{{0.03}}		&		\underline{\textbf{0.00}}		\\
MatrixFactor 400x1000			&		\underline{\textbf{0.00}}		&		\underline{\textbf{0.00}}		&		\underline{\textbf{0.00}}		&		\underline{\textbf{0.00}}		&		\underline{\textbf{0.00}}		\\
MatrixFactor 600x1000			&		\underline{\textbf{0.00}}		&		\underline{\textbf{0.00}}		&		{{0.12}}		&		\underline{\textbf{0.00}}		&		\underline{\textbf{0.00}}		\\
MatrixFactor 800x1000			&		{{0.43}}		&		{{0.01}}		&		\underline{\textbf{0.00}}		&		\underline{\textbf{0.00}}		&		\underline{\textbf{0.00}}		\\
MatrixFactor 1000x1000			&		{{0.09}}		&		\underline{\textbf{0.00}}		&		{{0.10}}		&		\underline{\textbf{0.00}}		&		{{0.03}}		\\[1ex]
\midrule																							
Average			&		0.71		&		0.34		&		0.47		&		0.22		&		\textbf{0.20}		\\
Max			&		4.51		&		2.65		&		3.19		&		1.49		&		\textbf{1.48}		\\
\bottomrule
\end{tabular}
\caption{
	Evaluation of metaheuristics on Medium Instances, 10~sec per run.
    Reported are the gaps, as percentages, to the best known solutions.
    Best value in a row is bold, and where heuristic finds the best known (objective value) solution, the gap is underlined.  (Note that due to rounding, a gap value of $0.00$ is not automatically the same as having found the best known.)
    }
\label{tab:internal-evaluation-medium}
\end{table*}

\begin{table*}[htb]
\centering
\begin{tabular}{@{} lrrrrr @{}}
\toprule
Instance & VNS & \opprob{} & MCHH & \reduced{} & \krow{2} \\

\midrule																	Rand 1000x5000			&		{{0.07}}	&		\underline{\textbf{0.00}}	&		{{0.08}}	&		{{0.04}}	&		{{0.04}}	\\
Rand 2000x5000			&		{{0.38}}	&		{{0.17}}	&		{{0.15}}	&		{{0.13}}	&		{\textbf{0.07}}	\\
Rand 3000x5000			&		{{0.50}}	&		{{0.19}}	&		{{0.22}}	&		{{0.24}}	&		{\textbf{0.12}}	\\
Rand 4000x5000			&		{{0.29}}	&		{{0.13}}	&		{{0.19}}	&		{{0.08}}	&		{\textbf{0.07}}	\\
Rand 5000x5000			&		{{0.38}}	&		{{0.31}}	&		{{0.31}}	&		{{0.23}}	&		{\textbf{0.11}}	\\[1ex]
																		
Biclique 1000x5000			&		{{0.92}}	&		{\textbf{0.06}}	&		{{0.23}}	&		{{0.22}}	&		{{0.08}}	\\
Biclique 2000x5000			&		{\textbf{0.05}}	&		{{0.37}}	&		{{0.53}}	&		{{0.57}}	&		{{0.52}}	\\
Biclique 3000x5000			&		\underline{\textbf{0.00}}	&		{{0.11}}	&		{{0.13}}	&		{{0.07}}	&		{{0.43}}	\\
Biclique 4000x5000			&		\underline{\textbf{0.00}}	&		\underline{\textbf{0.00}}	&		{{0.26}}	&		{{0.27}}	&		{{0.38}}	\\
Biclique 5000x5000			&		\underline{\textbf{0.00}}	&		{{0.16}}	&		\underline{\textbf{0.00}}	&		{{0.03}}	&		\underline{\textbf{0.00}}	\\[1ex]
																		
MaxInduced 1000x5000			&		{{0.21}}	&		{{0.01}}	&		{{0.01}}	&		{{0.05}}	&		{\textbf{0.01}}	\\
MaxInduced 2000x5000			&		{{0.36}}	&		{{0.08}}	&		{{0.19}}	&		{{0.01}}	&		{\textbf{0.01}}	\\
MaxInduced 3000x5000			&		{{0.53}}	&		{{0.11}}	&		{{0.21}}	&		{{0.20}}	&		{\textbf{0.08}}	\\
MaxInduced 4000x5000			&		{{0.52}}	&		{{0.30}}	&		{{0.28}}	&		{\textbf{0.14}}	&		{{0.20}}	\\
MaxInduced 5000x5000			&		{{0.52}}	&		{{0.32}}	&		{{0.42}}	&		{{0.23}}	&		{\textbf{0.14}}	\\[1ex]
																		
BMaxCut 1000x5000			&		{{2.57}}	&		{\textbf{0.71}}	&		{{1.39}}	&		{{2.90}}	&		{{2.69}}	\\
BMaxCut 2000x5000			&		{{5.61}}	&		{\textbf{2.63}}	&		{{3.41}}	&		{{3.99}}	&		{{3.75}}	\\
BMaxCut 3000x5000			&		{{6.00}}	&		{{2.86}}	&		{{4.11}}	&		{{3.35}}	&		{\textbf{2.69}}	\\
BMaxCut 4000x5000			&		{{6.09}}	&		{{4.33}}	&		{{4.07}}	&		{{3.41}}	&		{\textbf{3.34}}	\\
BMaxCut 5000x5000			&		{{5.28}}	&		{{3.76}}	&		{{4.34}}	&		{{2.65}}	&		{\textbf{2.49}}	\\[1ex]
																		
MatrixFactor 1000x5000			&		{{0.09}}	&		{{0.35}}	&		{{0.10}}	&		{\textbf{0.04}}	&		{{0.07}}	\\
MatrixFactor 2000x5000			&		{{0.41}}	&		{{0.12}}	&		{\textbf{0.11}}	&		{{0.13}}	&		{{0.16}}	\\
MatrixFactor 3000x5000			&		{{0.55}}	&		{{0.17}}	&		{{0.43}}	&		{{0.24}}	&		{\textbf{0.16}}	\\
MatrixFactor 4000x5000			&		{{0.45}}	&		{{0.34}}	&		{{0.43}}	&		{{0.28}}	&		{\textbf{0.13}}	\\
MatrixFactor 5000x5000			&		{{0.42}}	&		{{0.38}}	&		{{0.40}}	&		{{0.38}}	&		{\textbf{0.16}}	\\[1ex]
\midrule																		
Average			&		1.29	&		0.72	&		0.88	&		0.80	&		\textbf{0.72}	\\
Max			&		6.09	&		4.33	&		4.34	&		3.99	&		\textbf{3.75}	\\
\bottomrule
\end{tabular}
\caption{
	Evaluation of metaheuristics on Large Instances, 100~sec per run.
    The format of the table is identical to that of Table~\ref{tab:internal-evaluation-medium}.
    }
\label{tab:internal-evaluation-large}
\end{table*}  

 So far we have only been testing the performance of the metaheuristics on the training instance set.
 In Tables~\ref{tab:internal-evaluation-medium} and~\ref{tab:internal-evaluation-large} we report their performance on benchmark instances, giving 10~seconds per Medium instance and 100~seconds per Large instance.
 For each instance and metaheuristic, we report the percentage gap, between the solution obtained by that metaheuristic and the best known objective value for that instance.
 The best known objective values are obtained by recording the best solutions produced in all our experiments, not necessarily only the experiments reported in this paper.
 The best known solutions will be available for download, and their objective values are reported in Tables~\ref{tab:comparison-medium} and~\ref{tab:comparison-large}.

 The results of the experiments on benchmark instances generally positively correlate with the configuration objective function (\ref{eq:tuning-objective}) reported in Table~\ref{tab:heuristics500}, except that \opprob{}\ shows performance better than MCHH, and is competing with \reduced{} on Large instances.
 This shows a common problem that the evaluation by short runs on small instances, as used for training, may not always perfectly correlate with the performance of the heuristic on real (or benchmark) instances~\cite{Hutter2007}.
 However, in our case, the main conclusions are unaffected by this.
 In particular, we still observe that \krow{2} outperforms other metaheuristics, including \reduced{}, hence proving usefulness of the \Repair{} component.
 Also \krow{2} clearly outperforms MCHH demonstrating that even a restricted version of the CMCS schema is more robust than the MCHH schema; recall that CMCS is an extension of MCHH with conditional transitions.
 
 We made the source code of \krow{2} publicly available\footnote{\url{http://csee.essex.ac.uk/staff/dkarap/?page=publications&key=CMCS-BBQP}}.
 The code is in C\# and was tested on Windows and Linux machines.
 We note here that CMCS is relevant to the Programming by Optimisation (PbO) concept~\cite{Hoos2012}.
 We made sure that our code complies with the ``PbO Level~3'' standard, i.e.\ ``the software-development process is structured and carried out in a way that seeks to provide design choices and alternatives in many performance- relevant components of a project.''~\cite{Hoos2012}.
 Our code is not compliant with ``PbO Level~4'' because some of the choices made (specifically, the internal parameters of individual components) were not designed to be tuned along with the CMCS matrices; for details of PbO see~\cite{Hoos2012}.

\subsection{Comparison to the State-of-the-art}
\label{sec:comparison}

 There have been two published high-performance metaheuristics for BBQP: \emph{Iterated Local Search} by Duarte et al.~\cite{Duarte2014} and \emph{Tabu Search} by Glover et al.~\cite{Glover2015}.
 Both papers agree that their approaches perform similarly; in fact, following a sign test, Duarte et al.~conclude that ``there are not significant differences between both procedures''.
 At first, we compare \krow{2} to Tabu Search for which we have detailed experimental results~\cite{Glover2015}.
 Then we also compare \krow{2} to ILS using approach adopted in~\cite{Duarte2014}.

 Tabu Search has two phases: (i) a classic tabu search based on a relatively small neighbourhood, which runs until it fails to improve the solution, and (ii) a polishing procedure, similar to ours, which repeats a sequence of hill climbing operators \OptY{}, \FlpX{}, \OptX{} and \FlpY{} until a local maximum is reached.\footnote{In~\cite{Glover2015}, a composite of \OptY{} and \FlpX{} is called Flip-$x$-Float-$y$, and a composite of \OptX{} and \FlpY{} is called Flip-$y$-Float-$x$.}
 The whole procedure is repeated as many times as the time allows.

 The experiments in~\cite{Glover2015} were conducted on the same benchmark instances, first introduced in~\cite{Karapetyan2012} and now described in Section~\ref{sec:testbed} of this paper.
 Each run of Tabu Search was given 1000 seconds for Medium instances ($n = 1000$) and 10000 seconds for Large instances ($n = 5000$).
 In Table~\ref{tab:comparison-medium} we report the performance results of \krow{2}, our best performing metaheuristic, on Medium instances with 1, 10, 100 and 1000 second time limits, and in Table~\ref{tab:comparison-large} on Large instances with 10, 100, 1000 and 10000 second time limits, and explicitly compare those results to the performance of Tabu Search and so implicitly compare to the results of Duarte et al.\ \cite{Duarte2014} that were not significantly different from Tabu.

\begin{table*}[htb]
\centering
\begin{tabular}{@{} lrrrrrr @{}}
\toprule
	& 
    &	\multicolumn{4}{c}{\krow{2}}
    &	\multicolumn{1}{c@{}}{Tabu Search} \\
\cmidrule(lr){3-6}
\cmidrule(l){7-7}
Instance 
	&	Best known
    &	1 sec.
    &	10 sec.
    &	100 sec.
    &	1000 sec.
    &	1000 sec. \\
\midrule
Rand 200x1000 &	612,947 &	\textbf{\underline{0.00}} &	\textbf{\underline{0.00}} &	\textbf{\underline{0.00}} &	\textbf{\underline{0.00}} &	\textbf{\underline{0.00}} \\
Rand 400x1000 &	951,950 &	{0.05} &	\textbf{\underline{0.00}} &	\textbf{\underline{0.00}} &	\textbf{\underline{0.00}} &	\textbf{\underline{0.00}} \\
Rand 600x1000 &	1,345,748 &	\textbf{\underline{0.00}} &	\textbf{\underline{0.00}} &	\textbf{\underline{0.00}} &	\textbf{\underline{0.00}} &	{0.00} \\
Rand 800x1000 &	1,604,925 &	{0.09} &	\textbf{0.00} &	\textbf{0.00} &	\textbf{\underline{0.00}} &	{0.01} \\
Rand 1000x1000 &	1,830,236 &	\textbf{0.04} &	\textbf{0.04} &	\textbf{0.02} &	\textbf{\underline{0.00}} &	{0.07} \\[1ex]
						
Biclique 200x1000 &	2,150,201 &	\textbf{\underline{0.00}} &	\textbf{\underline{0.00}} &	\textbf{\underline{0.00}} &	\textbf{\underline{0.00}} &	\textbf{\underline{0.00}} \\
Biclique 400x1000 &	4,051,884 &	{0.27} &	{0.09} &	\textbf{\underline{0.00}} &	\textbf{\underline{0.00}} &	\textbf{\underline{0.00}} \\
Biclique 600x1000 &	5,501,111 &	\textbf{0.59} &	{1.48} &	\textbf{0.47} &	\textbf{0.47} &	{0.65} \\
Biclique 800x1000 &	6,703,926 &	\textbf{0.68} &	\textbf{0.56} &	\textbf{0.04} &	\textbf{0.04} &	{0.79} \\
Biclique 1000x1000 &	8,680,142 &	\textbf{0.10} &	\textbf{0.35} &	\textbf{0.35} &	\textbf{0.11} &	{0.91} \\[1ex]
						
MaxInduced 200x1000 &	513,081 &	\textbf{\underline{0.00}} &	\textbf{\underline{0.00}} &	\textbf{\underline{0.00}} &	\textbf{\underline{0.00}} &	\textbf{\underline{0.00}} \\
MaxInduced 400x1000 &	777,028 &	{0.01} &	\textbf{\underline{0.00}} &	\textbf{\underline{0.00}} &	\textbf{\underline{0.00}} &	\textbf{\underline{0.00}} \\
MaxInduced 600x1000 &	973,711 &	\textbf{\underline{0.00}} &	\textbf{\underline{0.00}} &	\textbf{\underline{0.00}} &	\textbf{\underline{0.00}} &	\textbf{\underline{0.00}} \\
MaxInduced 800x1000 &	1,205,533 &	\textbf{0.01} &	\textbf{0.00} &	\textbf{\underline{0.00}} &	\textbf{\underline{0.00}} &	{0.07} \\
MaxInduced 1000x1000 &	1,415,622 &	\textbf{0.03} &	\textbf{0.03} &	\textbf{0.03} &	\textbf{0.01} &	{0.06} \\[1ex]
						
BMaxCut 200x1000 &	617,700 &	{1.59} &	\textbf{0.06} &	\textbf{\underline{0.00}} &	\textbf{\underline{0.00}} &	{0.14} \\
BMaxCut 400x1000 &	951,726 &	{1.34} &	\textbf{0.40} &	\textbf{\underline{0.00}} &	\textbf{\underline{0.00}} &	{1.13} \\
BMaxCut 600x1000 &	1,239,982 &	\textbf{1.83} &	\textbf{0.53} &	\textbf{0.53} &	\textbf{0.37} &	{2.00} \\
BMaxCut 800x1000 &	1,545,820 &	{1.74} &	\textbf{1.05} &	\textbf{0.08} &	\textbf{0.08} &	{1.66} \\
BMaxCut 1000x1000 &	1,816,688 &	\textbf{1.83} &	\textbf{0.46} &	\textbf{0.23} &	\textbf{0.23} &	{2.47} \\[1ex]
						
MatrixFactor 200x1000 &	6,283 &	{0.18} &	\textbf{\underline{0.00}} &	\textbf{\underline{0.00}} &	\textbf{\underline{0.00}} &	\textbf{\underline{0.00}} \\
MatrixFactor 400x1000 &	9,862 &	\textbf{\underline{0.00}} &	\textbf{\underline{0.00}} &	\textbf{\underline{0.00}} &	\textbf{\underline{0.00}} &	\textbf{\underline{0.00}} \\
MatrixFactor 600x1000 &	12,902 &	{0.05} &	\textbf{\underline{0.00}} &	\textbf{\underline{0.00}} &	\textbf{\underline{0.00}} &	{0.03} \\
MatrixFactor 800x1000 &	15,466 &	{0.49} &	\textbf{\underline{0.00}} &	\textbf{\underline{0.00}} &	\textbf{\underline{0.00}} &	{0.19} \\
MatrixFactor 1000x1000 &	18,813 &	\textbf{0.08} &	\textbf{0.03} &	\textbf{\underline{0.00}} &	\textbf{\underline{0.00}} &	{0.11} \\
\midrule						
Average &&		0.44&	\textbf{0.20}&	\textbf{0.07}&	\textbf{0.05}&	0.41\\
Max &&		\textbf{1.83}&	\textbf{1.48}&	\textbf{0.53}&	\textbf{0.47}&	2.47\\
\bottomrule
\end{tabular}
\caption{
	Empirical comparison of the \krow{2} and Tabu Search \cite{Glover2015} (which performs on average similarly to the method of \cite{Duarte2014}) on the Medium instances.
	Reported are the gaps to the best known solution, in percent.
	As in Tables~\ref{tab:internal-evaluation-medium} and~\ref{tab:internal-evaluation-large}, where the heuristic finds the best known (objective value) solution, the value ($0.00$) is underlined.
    Where \krow{2} finds a solution at least as good as the one found by Tabu Search, the gap is shown in bold.
    Similarly, where Tabu Search (1000s) finds a solution at least as good as the one found by \krow{2} (1000s), the gap is shown in bold.
}
\label{tab:comparison-medium}
\end{table*}

\begin{table*}[htb]
\centering
\begin{tabular}{@{} lrrrrrr @{}}
\toprule
	& 
    &	\multicolumn{4}{c}{\krow{2}}
    &	\multicolumn{1}{c@{}}{Tabu Search} \\
\cmidrule(lr){3-6}
\cmidrule(l){7-7}
Instance 
	&	Best known 
    &	10 sec. 
    &	100 sec. 
    &	1000 sec. 
    &	10000 sec. 
    &	10000 sec. \\
\midrule
Rand 1000x5000 &	7,183,221 &	{0.04} &	{0.04} &	{0.01} &	{0.01} &	\textbf{0.01} \\
Rand 2000x5000 &	11,098,093 &	{0.18} &	\textbf{0.07} &	\textbf{0.07} &	\textbf{0.02} &	{0.09} \\
Rand 3000x5000 &	14,435,941 &	\textbf{0.16} &	\textbf{0.12} &	\textbf{0.11} &	\textbf{0.07} &	{0.22} \\
Rand 4000x5000 &	18,069,396 &	\textbf{0.14} &	\textbf{0.07} &	\textbf{0.01} &	\textbf{0.01} &	{0.19} \\
Rand 5000x5000 &	20,999,474 &	{0.26} &	\textbf{0.11} &	\textbf{0.08} &	\textbf{0.07} &	{0.25} \\[1ex]
						
Biclique 1000x5000 &	38,495,688 &	{0.22} &	{0.08} &	\textbf{0.02} &	\textbf{\underline{0.00}} &	{0.02} \\
Biclique 2000x5000 &	64,731,072 &	{1.67} &	\textbf{0.52} &	\textbf{0.19} &	\textbf{0.28} &	{0.94} \\
Biclique 3000x5000 &	98,204,538 &	{1.68} &	\textbf{0.43} &	\textbf{0.01} &	\textbf{0.04} &	{1.50} \\
Biclique 4000x5000 &	128,500,727 &	\textbf{0.38} &	\textbf{0.38} &	\textbf{0.22} &	\textbf{\underline{0.00}} &	{2.19} \\
Biclique 5000x5000 &	163,628,686 &	\textbf{0.38} &	\textbf{\underline{0.00}} &	\textbf{\underline{0.00}} &	\textbf{\underline{0.00}} &	{1.01} \\[1ex]
						
MaxInduced 1000x5000 &	5,465,051 &	\textbf{0.01} &	\textbf{0.01} &	\textbf{\underline{0.00}} &	\textbf{\underline{0.00}} &	{0.02} \\
MaxInduced 2000x5000 &	8,266,136 &	\textbf{0.10} &	\textbf{0.01} &	\textbf{0.01} &	\textbf{\underline{0.00}} &	{0.12} \\
MaxInduced 3000x5000 &	11,090,573 &	\textbf{0.15} &	\textbf{0.08} &	\textbf{0.04} &	\textbf{0.03} &	{0.18} \\
MaxInduced 4000x5000 &	13,496,469 &	\textbf{0.29} &	\textbf{0.20} &	\textbf{0.06} &	\textbf{0.05} &	{0.36} \\
MaxInduced 5000x5000 &	16,021,337 &	\textbf{0.19} &	\textbf{0.14} &	\textbf{0.08} &	\textbf{0.08} &	{0.29} \\[1ex]
						
BMaxCut 1000x5000 &	6,644,232 &	{2.98} &	{2.69} &	{2.17} &	\textbf{1.20} &	{1.70} \\
BMaxCut 2000x5000 &	10,352,878 &	{5.39} &	{3.75} &	{3.39} &	\textbf{1.80} &	{2.58} \\
BMaxCut 3000x5000 &	13,988,920 &	{3.49} &	\textbf{2.69} &	\textbf{1.99} &	\textbf{1.81} &	{3.45} \\
BMaxCut 4000x5000 &	17,090,794 &	{4.36} &	\textbf{3.34} &	\textbf{3.31} &	\textbf{2.31} &	{4.28} \\
BMaxCut 5000x5000 &	20,134,370 &	\textbf{3.15} &	\textbf{2.49} &	\textbf{2.34} &	\textbf{1.79} &	{3.90} \\[1ex]
						
MatrixFactor 1000x5000 &	71,485 &	{0.11} &	{0.07} &	\textbf{0.00} &	\textbf{0.00} &	{0.02} \\
MatrixFactor 2000x5000 &	108,039 &	{0.19} &	{0.16} &	\textbf{0.06} &	\textbf{0.04} &	{0.09} \\
MatrixFactor 3000x5000 &	144,255 &	\textbf{0.17} &	\textbf{0.16} &	\textbf{0.14} &	\textbf{0.11} &	{0.26} \\
MatrixFactor 4000x5000 &	179,493 &	\textbf{0.26} &	\textbf{0.13} &	\textbf{0.10} &	\textbf{0.10} &	{0.29} \\
MatrixFactor 5000x5000 &	211,088 &	\textbf{0.21} &	\textbf{0.16} &	\textbf{0.13} &	\textbf{0.04} &	{0.33} \\
\midrule						
Average &&		1.05&	\textbf{0.72}&	\textbf{0.58}&	\textbf{0.39}&	0.97\\
Max &&		5.39&	\textbf{3.75}&	\textbf{3.39}&	\textbf{2.31}&	4.28\\
\bottomrule
\end{tabular}
\caption{
	Empirical comparison of \krow{2} with Tabu Search~\cite{Glover2015} (which performs on average similarly to the method of \cite{Duarte2014}) on the Large instances. 
    The format of the table is identical to that of Table~\ref{tab:comparison-medium}. 
    }
\label{tab:comparison-large}
\end{table*}

 Given the same time, \krow{2} produces same (for 10 instances) or better (for 20 instances) solutions.
 The worst gap between best known and obtained solution (reported in the Max row at the bottom of each table) is also much larger for Tabu Search than for \krow{2}.
 \krow{2} clearly outperforms Tabu Search even if given a factor of 100 less time, and competes with it even if given a factor of 1000 less time.
 Thus we conclude that \krow{2} is faster than Tabu Search by two to three orders of magnitude.
 Further, we observe that \krow{2} does not converge prematurely, that is, it continues to improve the solution when given more time.
 
 As pointed out above, it is known from the literature that ILS~\cite{Duarte2014} performs similarly to Tabu Search, and hence the conclusions of the comparison between \krow{2} and Tabu Search can be extended to ILS as well.
 However, to verify this, we reproduced the experiment from \cite{Duarte2014}.
 In that experiment, Duarte et al.\ solved each of the medium and large instances, giving ILS 1000 seconds per run, and then reported the average objective value.
 We tested \krow{2} is exactly the same way, except that we allowed only 10 seconds per run.
 Despite a much lower time budget, our result of 14,523,968.32 is superior to the result of 14,455,832.30 reported in \cite[Table 8]{Duarte2014}.
 This direct experiment confirms that \krow{2} significantly outperforms ILS.

 We note here that this result is achieved in spite of \krow{2} consisting of simple components combined in an entirely automated way; without any human intelligence put into the detailed metaheuristic design.
 Instead, only a modest computational power (a few hours of CPU time) was required to obtain it.
 (Note that this computational power should not be compared to the running time of the algorithm itself; it is a replacement of expensive time of a human expert working on manual design of a high-performance solution method.)
 We believe that these results strongly support the idea of automated metaheuristic in general and CMCS schema in particular.

\tempclearpage
\section{Conclusions}
\label{sec:conclusion}

 In this work, we considered an important combinatorial optimisation problem called Bipartite Boolean Quadratic Programming Problem (BBQP)\@.
 We defined several algorithmic components for BBQP, primarily aiming at components for metaheuristics.
 To test and analyse the performance of the components, and to combine them in a powerful metaheuristic, we designed a flexible metaheuristic schema, which we call Conditional Markov Chain Search (CMCS), the behaviour of which is entirely defined by an explicit set of parameters and thus which is convenient for automated configuration.
 CMCS is a powerful schema with special cases covering several standard metaheuristics.
 Hence, to evaluate the performance of a metaheuristic on a specific problem class, we can configure the CMCS restricted to that metaheuristic, obtaining a nearly best possible metaheuristic of that particular type for that specific problem class.
 The key advantages of this approach include avoidance of human/expert bias in analysis of the components and metaheuristics, and complete automation of the typically time-consuming process of metaheuristic design.

 Of the methods we consider, the CMCS schema is potentially the most powerful as it includes the others as special cases, however, it has a lot of parameters, and this complicates the selection of the matrices. 
 To combat this, we proposed a special case of CMCS, \krow{$k$}, which is significantly easier to configure, but that still preserves much of the flexibility of the approach.

 By configuring several special cases of CMCS on a set of small instances and then testing them on benchmark instances, we learnt several lessons.
 In particular, we found out that CMCS schema, even if restricted to the \krow{2} schema, is significantly more powerful than VNS, \opprob{}\ and even MCHH (with a static transition matrix).
 We also verified that the new BBQP component, \Repair{}, is useful, as its inclusion in the pool of components improved the performance of \krow{2}.
 Finally, we showed that the best found strategies are often much more sophisticated than the strategies implemented in standard approaches.

 Our best performing metaheuristic, \krow{2}, clearly outperforms the previous state-of-the-art BBQP methods.
 Following a series of computational experiments, we estimated that \krow{2} is faster than those methods by roughly two to three orders of magnitude.

 \subsection{Future Work}

 A few other BBQP algorithmic components could be studied and exploited using the CMCS schema.
 Variations of the \Repair{} heuristic, as discussed in Section~\ref{sec:repair}, should be considered more thoroughly.
 Another possibility for creating a new class of powerful components is to reduce the entire problem by adding constraints of the form $x_i = x_{i'}$, $x_i \neq x_{i'}$ or $x_i = 1$, or even more sophisticated such as $x_i = x_{i'} \vee x_{i''}$.
 Note that such constraints effectively reduce the original problem to a smaller BBQP; then this smaller BBQP can be solved exactly or heuristically.
 Also note that if such constraints are generated to be consistent with the current solution then this approach can be used as a hill climbing operator.
 
 It is interesting to note that the reduced size subproblem could itself be solved using a version of CMCS configured to be effective for brief intense runs. 
 This gives the intriguing possibility of an upper-level CMCS in which one of the components uses a different CMCS -- though we expect that tuning such a system could be a significant, but interesting, challenge.
 
 The CMCS schema should be developed in several directions.
 First of all, it should be tested on other domains.
 Then a few extensions can be studied, e.g.\ one could add a ``termination'' component that would stop the search -- to allow variable running times.
 It is possible to add some form of memory and/or backtracking functionality, for example to implement a tabu-like mechanism. 
 Another direction of research is population-based extensions of CMCS\@.
 Of interest are efficient configuration procedures that would allow to include more components. 
 Finally, of course, one can study methods for online learning, that is adaptation of the transition probabilities during the search process itself; in which case it would be most natural to call the method ``Conditional Markov Chain Hyper-heuristic.''
 
\section*{Acknowledgement}

 This research work was partially supported by an NSERC Discovery accelerator supplement awarded to Abraham P. Punnen, EPSRC grants EP/H000968/1 and EP/F033214/1 (``The LANCS Initiative''), and also LANCS Initiative International Scientific Outreach Fund which supported the visit of Daniel Karapetyan to the Simon Fraser University.



\bibliographystyle{model2-names}
\bibliography{bqp}{}

\begin{thebibliography}{28}
\expandafter\ifx\csname natexlab\endcsname\relax\def\natexlab#1{#1}\fi
\providecommand{\url}[1]{\texttt{#1}}
\providecommand{\href}[2]{#2}
\providecommand{\path}[1]{#1}
\providecommand{\DOIprefix}{doi:}
\providecommand{\ArXivprefix}{arXiv:}
\providecommand{\URLprefix}{URL: }
\providecommand{\Pubmedprefix}{pmid:}
\providecommand{\doi}[1]{\href{http://dx.doi.org/#1}{\path{#1}}}
\providecommand{\Pubmed}[1]{\href{pmid:#1}{\path{#1}}}
\providecommand{\bibinfo}[2]{#2}
\ifx\xfnm\relax \def\xfnm[#1]{\unskip,\space#1}\fi
\bibitem[{Adenso-D\'iaz and Laguna(2006)}]{Belarmino2006}
\bibinfo{author}{Adenso-D\'iaz, B.}, \bibinfo{author}{Laguna, M.},
  \bibinfo{year}{2006}.
\newblock \bibinfo{title}{Fine-tuning of algorithms using fractional
  experimental designs and local search}.
\newblock \bibinfo{journal}{Operations Research} \bibinfo{volume}{54},
  \bibinfo{pages}{99--114}.
\bibitem[{Ahuja et~al.(2002)Ahuja, Ergun, Orlin and Punnen}]{Ajuja2002}
\bibinfo{author}{Ahuja, R.K.}, \bibinfo{author}{Ergun, {\"O}.},
  \bibinfo{author}{Orlin, J.B.}, \bibinfo{author}{Punnen, A.P.},
  \bibinfo{year}{2002}.
\newblock \bibinfo{title}{A survey of very large-scale neighborhood search
  techniques}.
\newblock \bibinfo{journal}{Discrete Applied Mathematics}
  \bibinfo{volume}{123}, \bibinfo{pages}{75--102}.
\bibitem[{Alon and Naor(2006)}]{Alon2006}
\bibinfo{author}{Alon, N.}, \bibinfo{author}{Naor, A.}, \bibinfo{year}{2006}.
\newblock \bibinfo{title}{{Approximating the Cut-Norm via Grothendieck's
  Inequality}}.
\newblock \bibinfo{journal}{SIAM Journal on Computing} \bibinfo{volume}{35},
  \bibinfo{pages}{787--803}.
\bibitem[{Amb\"{u}hl et~al.(2011)Amb\"{u}hl, Mastrolilli and
  Svensson}]{Ambuhl2011}
\bibinfo{author}{Amb\"{u}hl, C.}, \bibinfo{author}{Mastrolilli, M.},
  \bibinfo{author}{Svensson, O.}, \bibinfo{year}{2011}.
\newblock \bibinfo{title}{Inapproximability results for maximum edge biclique,
  minimum linear arrangement, and sparsest cut}.
\newblock \bibinfo{journal}{SIAM Journal on Computing} \bibinfo{volume}{40},
  \bibinfo{pages}{567--596}.
\bibitem[{Bezerra et~al.(2015)Bezerra, L{\'o}pez-Ib{\'a}{\~{n}}ez and
  St{\"u}tzle}]{BezerraEtal2015:component-MOO}
\bibinfo{author}{Bezerra, L.C.T.}, \bibinfo{author}{L{\'o}pez-Ib{\'a}{\~{n}}ez,
  M.}, \bibinfo{author}{St{\"u}tzle, T.}, \bibinfo{year}{2015}.
\newblock \bibinfo{title}{Evolutionary Multi-Criterion Optimization: 8th
  International Conference, EMO 2015, Guimar{\~a}es, Portugal, March 29 --April
  1, 2015. Proceedings, Part I}. \bibinfo{publisher}{Springer International
  Publishing}, \bibinfo{address}{Cham}. chapter \bibinfo{chapter}{Comparing
  Decomposition-Based and Automatically Component-Wise Designed Multi-Objective
  Evolutionary Algorithms}.
\newblock pp. \bibinfo{pages}{396--410}.
\newblock \URLprefix \url{http://dx.doi.org/10.1007/978-3-319-15934-8_27},
  \DOIprefix\doi{10.1007/978-3-319-15934-8_27}.
\bibitem[{Billionnet(2004)}]{Billionnet2004}
\bibinfo{author}{Billionnet, A.}, \bibinfo{year}{2004}.
\newblock \bibinfo{title}{{Quadratic 0-1 bibliography}} \URLprefix
  \url{http://cedric.cnam.fr/fichiers/RC611.pdf}.
\bibitem[{Chang et~al.(2012)Chang, Vakati, Krause and Eulenstein}]{Chang2012}
\bibinfo{author}{Chang, W.C.}, \bibinfo{author}{Vakati, S.},
  \bibinfo{author}{Krause, R.}, \bibinfo{author}{Eulenstein, O.},
  \bibinfo{year}{2012}.
\newblock \bibinfo{title}{{Exploring biological interaction networks with
  tailored weighted quasi-bicliques.}}
\newblock \bibinfo{journal}{BMC bioinformatics} \bibinfo{volume}{13 Suppl 1},
  \bibinfo{pages}{S16}.
\bibitem[{Cowling et~al.(2001)Cowling, Kendall and Soubeiga}]{Cowling2001}
\bibinfo{author}{Cowling, P.}, \bibinfo{author}{Kendall, G.},
  \bibinfo{author}{Soubeiga, E.}, \bibinfo{year}{2001}.
\newblock \bibinfo{title}{A hyperheuristic approach to scheduling a sales
  summit}, in: \bibinfo{editor}{Burke, E.}, \bibinfo{editor}{Erben, W.} (Eds.),
  \bibinfo{booktitle}{Selected papers from the 3rd International Conference on
  the Practice and Theory of Automated Timetabling (PATAT 2001)},
  \bibinfo{publisher}{Springer}. pp. \bibinfo{pages}{176--190}.
\newblock \DOIprefix\doi{10.1007/3-540-44629-X_11}.
\bibitem[{Duarte et~al.(2014)Duarte, Laguna, Mart\'i and
  S\'anchez-Oro}]{Duarte2014}
\bibinfo{author}{Duarte, A.}, \bibinfo{author}{Laguna, M.},
  \bibinfo{author}{Mart\'i, R.}, \bibinfo{author}{S\'anchez-Oro, J.},
  \bibinfo{year}{2014}.
\newblock \bibinfo{title}{Optimization procedures for the bipartite
  unconstrained 0-1 quadratic programming problem}.
\newblock \bibinfo{journal}{Computers \& Operations Research}
  \bibinfo{volume}{51}, \bibinfo{pages}{123--129}.
\bibitem[{Gillis and Glineur(2011)}]{Gillis2011}
\bibinfo{author}{Gillis, N.}, \bibinfo{author}{Glineur, F.},
  \bibinfo{year}{2011}.
\newblock \bibinfo{title}{Low-rank matrix approximation with weights or missing
  data is {NP}-hard}.
\newblock \bibinfo{journal}{SIAM Journal on Matrix Analysis and Applications}
  \bibinfo{volume}{32}, \bibinfo{pages}{1149--1165}.
\bibitem[{Glover et~al.(2015)Glover, Ye, Punnen and Kochenberger}]{Glover2015}
\bibinfo{author}{Glover, F.}, \bibinfo{author}{Ye, T.},
  \bibinfo{author}{Punnen, A.}, \bibinfo{author}{Kochenberger, G.},
  \bibinfo{year}{2015}.
\newblock \bibinfo{title}{Integrating tabu search and {VLSN} search to develop
  enhanced algorithms: A case study using bipartite boolean quadratic
  programs}.
\newblock \bibinfo{journal}{European Journal of Operational Research}
  \bibinfo{volume}{241}, \bibinfo{pages}{697--707}.
\bibitem[{Hoos(2012)}]{Hoos2012}
\bibinfo{author}{Hoos, H.H.}, \bibinfo{year}{2012}.
\newblock \bibinfo{title}{Programming by optimization}.
\newblock \bibinfo{journal}{Communications of the {ACM}} \bibinfo{volume}{55},
  \bibinfo{pages}{70--80}.
\bibitem[{Hutter et~al.(2009)Hutter, Hoos, Leyton-Brown and
  St{\"u}tzle}]{HutterEtal2009:ParamILS}
\bibinfo{author}{Hutter, F.}, \bibinfo{author}{Hoos, H.H.},
  \bibinfo{author}{Leyton-Brown, K.}, \bibinfo{author}{St{\"u}tzle, T.},
  \bibinfo{year}{2009}.
\newblock \bibinfo{title}{Param{ILS}: An automatic algorithm configuration
  framework}.
\newblock \bibinfo{journal}{Journal of Artificial Research}
  \bibinfo{volume}{36}, \bibinfo{pages}{267--306}.
\newblock \DOIprefix\doi{10.1613/jair.2861}.
\bibitem[{Hutter et~al.(2007)Hutter, Hoos and St\"{u}tzle}]{Hutter2007}
\bibinfo{author}{Hutter, F.}, \bibinfo{author}{Hoos, H.H.},
  \bibinfo{author}{St\"{u}tzle, T.}, \bibinfo{year}{2007}.
\newblock \bibinfo{title}{Automatic algorithm configuration based on local
  search}, in: \bibinfo{booktitle}{Proceedings of the 22nd National Conference
  on Artificial Intelligence - Volume 2}, \bibinfo{publisher}{AAAI Press}. pp.
  \bibinfo{pages}{1152--1157}.
\bibitem[{Karapetyan and Punnen(2012)}]{Karapetyan2012}
\bibinfo{author}{Karapetyan, D.}, \bibinfo{author}{Punnen, A.P.},
  \bibinfo{year}{2012}.
\newblock \bibinfo{title}{Heuristic algorithms for the bipartite unconstrained
  0-1 quadratic programming problem} \URLprefix
  \url{http://arxiv.org/abs/1210.3684}.
\bibitem[{Koyut\"{u}rk et~al.(2005)Koyut\"{u}rk, Grama and
  Ramakrishnan}]{Koyuturk2005}
\bibinfo{author}{Koyut\"{u}rk, M.}, \bibinfo{author}{Grama, A.},
  \bibinfo{author}{Ramakrishnan, N.}, \bibinfo{year}{2005}.
\newblock \bibinfo{title}{{Compression, clustering, and pattern discovery in
  very high-dimensional discrete-attribute data sets}}.
\newblock \bibinfo{journal}{IEEE Transactions on Knowledge and Data
  Engineering} \bibinfo{volume}{17}, \bibinfo{pages}{447--461}.
\bibitem[{Koyut\"{u}rk et~al.(2006)Koyut\"{u}rk, Grama and
  Ramakrishnan}]{Koyuturk2006}
\bibinfo{author}{Koyut\"{u}rk, M.}, \bibinfo{author}{Grama, A.},
  \bibinfo{author}{Ramakrishnan, N.}, \bibinfo{year}{2006}.
\newblock \bibinfo{title}{{Nonorthogonal decomposition of binary matrices for
  bounded-error data compression and analysis}}.
\newblock \bibinfo{journal}{ACM Transactions on Mathematical Software}
  \bibinfo{volume}{32}, \bibinfo{pages}{33--69}.
\bibitem[{Louren\c{c}o et~al.(2010)Louren\c{c}o, Martin and
  St\"{u}tzle}]{Lourenco2010}
\bibinfo{author}{Louren\c{c}o, H.R.}, \bibinfo{author}{Martin, O.C.},
  \bibinfo{author}{St\"{u}tzle, T.}, \bibinfo{year}{2010}.
\newblock \bibinfo{title}{Iterated local search: Framework and applications
  handbook of metaheuristics}, \bibinfo{publisher}{Springer US},
  \bibinfo{address}{Boston, MA}. volume \bibinfo{volume}{146} of
  \textit{\bibinfo{series}{International Series in Operations Research \&
  Management Science}}. chapter~\bibinfo{chapter}{12}, pp.
  \bibinfo{pages}{363--397}.
\newblock \DOIprefix\doi{10.1007/978-1-4419-1665-5\_12}.
\bibitem[{Lu et~al.(2011)Lu, Vaidya, Atluri, Shin and Jiang}]{Lu2011}
\bibinfo{author}{Lu, H.}, \bibinfo{author}{Vaidya, J.},
  \bibinfo{author}{Atluri, V.}, \bibinfo{author}{Shin, H.},
  \bibinfo{author}{Jiang, L.}, \bibinfo{year}{2011}.
\newblock \bibinfo{title}{Weighted rank-one binary matrix factorization}, in:
  \bibinfo{booktitle}{Proceedings of the Eleventh SIAM International Conference
  on Data Mining}, \bibinfo{publisher}{SIAM / Omnipress}. pp.
  \bibinfo{pages}{283--294}.
\bibitem[{McClymont and
  Keedwell(2011)}]{McClymontKeedwell:GECCO2011:MCHH-selective-HH}
\bibinfo{author}{McClymont, K.}, \bibinfo{author}{Keedwell, E.C.},
  \bibinfo{year}{2011}.
\newblock \bibinfo{title}{Markov chain hyper-heuristic ({MCHH}): An online
  selective hyper-heuristic for multi-objective continuous problems}, in:
  \bibinfo{booktitle}{Proceedings of the 13th Annual Conference on Genetic and
  Evolutionary Computation}, \bibinfo{publisher}{ACM}, \bibinfo{address}{New
  York, NY, USA}. pp. \bibinfo{pages}{2003--2010}.
\newblock \DOIprefix\doi{10.1145/2001576.2001845}.
\bibitem[{Papadimitriou(1991)}]{Papadimitriou1991:selecting-truth-assignment}
\bibinfo{author}{Papadimitriou, C.H.}, \bibinfo{year}{1991}.
\newblock \bibinfo{title}{On selecting a satisfying truth assignment}, in:
  \bibinfo{booktitle}{Foundations of Computer Science, 1991. Proceedings., 32nd
  Annual Symposium on}, pp. \bibinfo{pages}{163--169}.
\newblock \DOIprefix\doi{10.1109/SFCS.1991.185365}.
\bibitem[{Punnen et~al.(2015a)Punnen, Sripratak and Karapetyan}]{Punnen2015}
\bibinfo{author}{Punnen, A.P.}, \bibinfo{author}{Sripratak, P.},
  \bibinfo{author}{Karapetyan, D.}, \bibinfo{year}{2015}a.
\newblock \bibinfo{title}{Average value of solutions for the bipartite boolean
  quadratic programs and rounding algorithms}.
\newblock \bibinfo{journal}{Theoretical Computer Science}
  \bibinfo{volume}{565}, \bibinfo{pages}{77--89}.
\bibitem[{Punnen et~al.(2015b)Punnen, Sripratak and Karapetyan}]{Punnen2012}
\bibinfo{author}{Punnen, A.P.}, \bibinfo{author}{Sripratak, P.},
  \bibinfo{author}{Karapetyan, D.}, \bibinfo{year}{2015}b.
\newblock \bibinfo{title}{The bipartite unconstrained 0-1 quadratic programming
  problem: polynomially solvable cases}.
\newblock \bibinfo{journal}{Discrete Applied Mathematics}
  \bibinfo{volume}{193}, \bibinfo{pages}{1--10}.
\bibitem[{Selman et~al.(1995)Selman, Kautz and Cohen}]{Selman95localsearch}
\bibinfo{author}{Selman, B.}, \bibinfo{author}{Kautz, H.},
  \bibinfo{author}{Cohen, B.}, \bibinfo{year}{1995}.
\newblock \bibinfo{title}{Local search strategies for satisfiability testing},
  in: \bibinfo{booktitle}{DIMACS series in discrete mathematics and theoretical
  computer science}, pp. \bibinfo{pages}{521--532}.
\bibitem[{Shen et~al.(2009)Shen, Ji and Ye}]{Shen2009}
\bibinfo{author}{Shen, B.h.}, \bibinfo{author}{Ji, S.}, \bibinfo{author}{Ye,
  J.}, \bibinfo{year}{2009}.
\newblock \bibinfo{title}{{Mining discrete patterns via binary matrix
  factorization}}, in: \bibinfo{booktitle}{Proceedings of the 15th ACM SIGKDD
  international conference on Knowledge discovery and data mining},
  \bibinfo{publisher}{ACM Press}. pp. \bibinfo{pages}{757--766}.
\bibitem[{Tan(2008)}]{Tan2008}
\bibinfo{author}{Tan, J.}, \bibinfo{year}{2008}.
\newblock \bibinfo{title}{{Inapproximability of Maximum Weighted Edge Biclique
  and Its Applications}}, in: \bibinfo{booktitle}{Proceedings of the 5th
  international conference on Theory and applications of models of
  computation}, \bibinfo{publisher}{Springer-Verlag}. pp.
  \bibinfo{pages}{282--293}.
\bibitem[{Tanay et~al.(2002)Tanay, Sharan and Shamir}]{Tanay2002}
\bibinfo{author}{Tanay, A.}, \bibinfo{author}{Sharan, R.},
  \bibinfo{author}{Shamir, R.}, \bibinfo{year}{2002}.
\newblock \bibinfo{title}{{Discovering statistically significant biclusters in
  gene expression data}}.
\newblock \bibinfo{journal}{Bioinformatics} \bibinfo{volume}{18},
  \bibinfo{pages}{S136--S144}.
\bibitem[{Zweben et~al.(1993)Zweben, Davis, Daun and
  Deale}]{ZwebenEtal1993:scheduling-with-iterative-repair}
\bibinfo{author}{Zweben, M.}, \bibinfo{author}{Davis, E.},
  \bibinfo{author}{Daun, B.}, \bibinfo{author}{Deale, M.J.},
  \bibinfo{year}{1993}.
\newblock \bibinfo{title}{Scheduling and rescheduling with iterative repair}.
\newblock \bibinfo{journal}{IEEE Transactions on Systems, Man, and Cybernetics}
  \bibinfo{volume}{23}, \bibinfo{pages}{1588--1596}.
\newblock \DOIprefix\doi{10.1109/21.257756}.

\end{thebibliography}

\end{document}